# A new method for faster and more accurate inference of species associations from big community data

Maximilian Pichler[1,*], Florian Hartig[1]

[1] Theoretical Ecology, University of Regensburg, Universitätsstraße 31, 93053 Regensburg, Germany

*corresponding author, maximilian.pichler@biologie.uni-regensburg.de

# Abstract

1. Joint Species Distribution models (JSDMs) explain spatial variation in community composition by contributions of the environment, biotic associations, and possibly spatially structured residual covariance. They show great promise as a general analytical framework for community ecology and macroecology, but current JSDMs, even when approximated by latent variables, scale poorly on large datasets, limiting their usefulness for currently emerging big (e.g., metabarcoding and metagenomics) community datasets.

2. Here, we present a novel, more scalable JSDM (sjSDM) that circumvents the need to use latent variables by using a Monte-Carlo integration of the joint JSDM likelihood and allows flexible elastic net regularization on all model components. We implemented sjSDM in PyTorch, a modern machine learning framework that can make use of CPU and GPU calculations. Using simulated communities with known species-species associations and different number of species and sites, we compare sjSDM with state-of-the-art JSDM implementations to determine computational runtimes and accuracy of the inferred species-species and species-environmental associations.

3. We find that sjSDM is orders of magnitude faster than existing JSDM algorithms (even when run on the CPU) and can be scaled to very large datasets. Despite the dramatically improved speed, sjSDM produces more accurate estimates of species association structures than alternative JSDM implementations. We demonstrate the applicability of sjSDM to big community data using eDNA case study with thousands of fungi operational taxonomic units (OTU).

4. Our sjSDM approach makes the analysis of JSDMs to large community datasets with hundreds or thousands of species possible, substantially extending the applicability of JSDMs in ecology. We provide our method in an R package to facilitate its applicability for practical data analysis.

# Introduction

Understanding the structure and assembly of ecological communities is a central concern for ecology, biogeography and macroecology (Vellend 2010). The question is tightly connected to important research programs of the field, including coexistence theory (see Chesson 2000; e.g. Levine *et al.* 2017), the emergence of diversity patterns (e.g. Pontarp *et al.* 2019) or understanding ecosystem responses to global change (Urban *et al.* 2016).

The statistical analysis of spatial community data is currently dominated by two major ecological frameworks: metacommunity theory (see Leibold *et al.* 2004) and species distribution models (SDMs, Elith & Leathwick 2009). Metacommunity theory formed in the last two decades as the study of the spatial processes that give rise to regional community assembly (e.g. Leibold *et al.* 2004; Leibold & Chase 2017). The current analytical frameworks of metacommunity theory are based on ordination techniques (e.g. Leibold & Mikkelson 2002) and statistical variation partitioning, which disentangles abiotic and spatial contributions to community assembly (see Cottenie 2005; Leibold & Chase 2017). SDMs are statistical models that link abiotic covariates to species occurrences. They are widely used in spatial ecology, for example to study invading species (Gallien *et al.* 2012; Mainali *et al.* 2015) or species responses to climate change (Thuiller *et al.* 2006).

A key limitation of both variation partitioning and SDMs, noted in countless studies, is that they do not account for species interactions. Both approaches essentially assume that species depend only on space and the environment (Cottenie 2005; Peres-Neto & Legendre 2010; Dormann *et al.* 2012; Wisz *et al.* 2013), whereas we know that in reality species can also influence each other through competition, predation, facilitation and other processes (Gilbert & Bennett 2010; Van der Putten *et al.* 2010; see Mittelbach & Schemske 2015; see Leibold & Chase 2017).

Joint Species Distribution Models (JSDM) recently emerged as a novel analytical framework promising to integrate species interactions into metacommunity and macroecology (Leibold *et al.* 2020). JSDMs are similar to SDMs in that they describe species occurrence as a function of the environment, but additionally consider the possibility of species-species associations. By an association, we mean that two species tend to appear together more or less often than expected from their environmental responses alone. Whether those association originate from biotic interactions (e.g. competition, predation, parasitism, mutualism) or other reasons (e.g. unmeasured environmental predictors) needs to be carefully considered (see Dormann *et al.* 2018; see Blanchet *et al.* 2020; König *et al.* 2021; Poggiato *et al.* 2021). Still, when appropriately interpreted, JSDMs combine the essential processes believed to be responsible for the assembly of ecological communities: environment, space, and biotic interactions; and they could be applied to large-scale as well as for regional metacommunity analyses (e.g. Gilbert & Bennett 2010; Mittelbach & Schemske 2015; Leibold & Chase 2017).

Recent interest in JSDMs was further fueled by the emergence of high-throughput technologies that are currently revolutionizing our capacities for observing community data (e.g. Pimm *et al.* 2015). We can now detect hundreds or even thousands of species from environmental DNA (eDNA) or bulk-sampled DNA (Cristescu 2014; Deiner *et al.* 2017; see also Bálint *et al.* 2018; Humphreys *et al.* 2019; Tikhonov *et al.* 2019a) in a given sample, and next generation sequencing (NGS) has become cheap enough that this process could be replicated at scale. Other emerging technologies will likely also produce large amounts of community data, such as automatic species recognition (Guirado *et al.* 2018; e.g. Tabak *et al.* 2019) from acoustic recordings. Recent studies have used these methods to generate community inventories of fish (see Desjonquères *et al.* 2019.; e.g. Picciulin *et al.* 2019), forest wildlife (e.g. see Wrege *et al.* 2017), bird communities (Fritzler *et al.* 2017; Lasseck 2018; Wood *et al.* 2019) or bats (e.g. Mac Aodha *et al.* 2018). Jointly, these developments mean that large spatial community datasets will become available in the near future, and ecologists have to consider how to best analyze them.

Joint species distribution models would seem the natural analytical approach for these emerging new data, given their ability to separate the processes essential for spatial community assembly. Current JSDM software, however, have severe limitations for processing such large (and/or wide) datasets. Early JSDMs were based on the multivariate probit model (MVP, Chib & Greenberg 1998), which describes species-species associations via a covariance matrix (e.g. Ovaskainen *et al.* 2010; Clark *et al.* 2014; Pollock *et al.* 2014; Golding *et al.* 2015; Hui 2016). The limitation of the MVP approach is that it scales poorly for species-rich data, as the number of parameters in the species-species covariance matrix increases quadratically with the number of species (see Warton *et al.* 2015).

The current solution to this problem are latent-variable models (LVM), which replace the covariance matrix with a small number of latent variables (see Warton *et al.* 2015). The LVM reparameterization makes the estimation of MVP models computationally more efficient (see Warton *et al.* 2015; Ovaskainen *et al.* 2017b; Tikhonov *et al.* 2017; Niku *et al.* 2019; Norberg *et al.* 2019; Tikhonov *et al.* 2019a). That, however, does not mean that simultaneously estimating species' abiotic preferences and species-species associations with LVM models is fast. Integrating out the latent variables requires MCMC sampling or numerical approximations (e.g. Laplace, variational inference, see Niku *et al.* 2019), which is computationally costly and can fail to converge. For communities with hundreds of species, computational runtimes of current LVMs can still exceed hours or days (e.g. Tikhonov *et al.* 2019a; Wilkinson *et al.* 2019). This poses severe limitations for analyzing eDNA data, which can include thousands of species or operational taxonomic units (OTU, e.g. Frøslev *et al.* 2019). Moreover, LVMs also scale disadvantageously with the number of sites, because each site introduces additional parameters in the latent variables (Skrondal & Rabe-Hesketh 2004; Bartholomew *et al.* 2011). Thus, the advantage of the LVM over the full-MVP model decreases with increasing numbers of sites (on the order of thousands). An important challenge for the field is therefore to make JSDMs fast enough for big datasets (Krapu & Borsuk 2020).

A second question for JSDM development is the accuracy of inferred species associations. Surprisingly little is known about this question. Most existing JSDM assessments (e.g. Norberg *et al.* 2019; e.g. Tobler *et al.* 2019; Wilkinson *et al.* 2019) concentrate on runtime, predictive performance, or on aggregated measures of accuracy that do not necessarily capture the error of the estimated species-association structure (but see Zurell *et al.* 2018). From a statistical perspective, it is clear that estimating a large species covariance matrix with limited data must have considerable error.

The LVM approach, however, not only makes the models faster, but also a reduces the number of free parameters (see Warton *et al.* 2015), which should theoretically reduce the variance (and thus the error) of the species-species covariance estimates, but possibly at the cost of a certain bias. The trade-off between bias and variance is controlled by the number of latent variables: when the number of latent variables is similar to the number of species, the LVM model will be as flexible (and unbiased) as the full-MVP model. The fewer latent variables are used, the stronger the reduction in variance and the potential increase in bias. In practice, the number of latent variables is usually chosen much smaller than the number of species (the highest value we saw was 32 with hundreds of species in Tikhonov *et al.* 2019a), which means that JSDMs fitted currently by LVMs could show biases due to the regularization (regularization is used to control the bias-variance tradeoff by reducing the variance and increasing the bias) induced by the LVM structure (Stein 2014).

While trading off some bias against a reduction in variance is fundamental to all regularization approaches, and no concern as such, it seems important to understand the nature of the bias that is created by the LVM structure and examine if alternative regularization structures are more appropriate. Similar to LVMs, spatial models for large data often use a low rank approximation of the covariance matrix (e.g. Stein 2007, 2014; e.g. Sang *et al.* 2011). For Gaussian process models, it has been shown that this approximation captures the overall structure well (in the sense that the magnitude of covariances is captured well), but at the costs of larger errors in local structures (see Stein 2014). We conjecture that LVMs with a small number of latent variables behave analogous – with a few latent variables, it will be difficult to model a specific covariance structure without unintentionally introducing other covariances elsewhere, but it could be possible to generate a good approximation of the overall correlation level between species.

Here, we propose a new method for estimating JSDMs, called scalable JSDM (sjSDM), that addresses many of the above-mentioned problems. By using a Monte-Carlo approach (originally proposed by Chen *et al.* 2018) that can be outsourced to graphical processing units (GPUs), sjSDM is able to fit JSDMs with a full covariance structure, without having to resort to latent variables, extremely fast. To address the issue of overfitting due to the increased number of parameters compared to state-of-the-art latent variable JSDMs, we introduce a new regularization approach, which directly targets the covariance matrix of the full-MVP model. Additionally, we propose a method for optimizing the regularization strength based on tuning the parameter under a k-fold cross-validation.

To demonstrate the beneficial properties of the new model structure, we assess: (i) its computational runtime on GPUs and CPUs (ii) the accuracy of inferred species-species associations and species' environmental responses and (iii) its predictive performance. We compare the performance of sjSDM to several state-of-the-art JSDM software packages ( Hmsc, gllvm, and BayesComm; see also Harris 2015; Clark *et al.* 2017; Vieilledent & Clément 2019), as well as results from a recent JSDM model comparison (Wilkinson *et al.* 2019). Finally, to illustrate the applicability of our approach to wide community data, we additionally applied our model to a community eDNA dataset containing 3,649 fungi OTU over 125 sites.

# Methods

## The structure of the JSDM problem

Species-environment associations are classically addressed by species distribution models (SDM), which estimate the expected probability of presence of a species as a function of the environmental predictors. The functional form of the niche can be expressed by GLMs, or by more flexible (i.e. nonlinear and/or non-parametric) approaches such as generalized additive models, boosted regression trees or Random Forest (Elith & Leathwick 2009; e.g. Ingram *et al.* 2020).

A JSDM generalizes this approach by including the possibility of residual species-species correlations (in the literature usually called species-species associations). The most common JSDM structure is the multivariate probit model (MVP), which describes the site by species matrix $Y_{ij}$ ($Y_{ij} = 1$ if species $j = 1,…,J$ is present at site $i = 1,…,I$ or $Y_{ij} = 0$ if species $j$ is absent) as a function of the environmental covariates $X_{in}$ ($n = 1,…,N$ covariates), and the covariance matrix (species associations) $\Sigma$ accounts for correlations in $e_{ij}$ :

$$Z_{ij} = \beta_{j0} + \sum_{n=1}^{N} X_{in} * \beta_{nj} + e_{ij}$$

$$Y_{ij} = 1(Z_{ij} > 0)$$

$$\boldsymbol{e}_i \sim MVN(0, \Sigma)$$

(1)

Following parameter inference, the fitted species-species covariance matrix $\Sigma$ is normally transformed into a correlation matrix for further interpretation.

# Current approaches to fit the JSDM model structure

The model structure described in Eq. 1 can be fitted directly using the probit link, and the first JSDMs used this approach (Chib & Greenberg 1998; Pollock *et al.* 2014; see Wilkinson *et al.* 2019). Fitting the full-MVP model directly, however, has two drawbacks: first, calculating likelihoods for large covariance matrices is computationally costly. Second, the number of parameters in the covariance matrix for *j* species increases quadratically as $j*(j-1)/2$, i.e., for 50 species there are 2250 parameters to fit.

Because of these problems, a series of papers (Warton *et al.* 2015; Ovaskainen *et al.* 2016) introduced the latent-variable model (see Skrondal & Rabe-Hesketh 2004) to the JSDM problem. The latent-variable JSDM approximates the species-species covariance by introducing a number of latent covariates (= latent variables), which act exactly like real environmental covariates, except that their values are estimated as well. Species that react (via their factor loadings) similarly or differently to the latent variables thus show positive or negative associations, respectively (see Warton *et al.* 2015; Ovaskainen *et al.* 2017; Wilkinson *et al.* 2019 for details). The factor loadings can be translated into a species-species covariance matrix: $\Sigma = \lambda * \lambda^T$ ($\lambda$ = matrix of factor loadings). The latent variables can be interpreted as unobserved environmental predictors, but they can also be viewed as a purely technical approach to regularized low-rank reparameterization of the covariance matrix. One advantage of the LVM is that the latent variables can be used for constrained (LVM with environmental predictors) and unconstrained ordination (LVM without environmental predictors) (Warton *et al.* 2015). The complexity of association structure can be set via the number of latent variables (usually to a low number, see Warton *et al.* 2015)

# An alternative approach to fit the JSDM structure

Because the latent-variable models still have computational limitations, and because of the need for flexible regularization discussed in the introduction, we propose a different approach to fit the model structure in Eq. 1. The full-MVP assumes that the observed binary occurrence vector $Y_i \in \{0,1\}^J$ arises as the sign of the latent gaussian variable $Y_i^* \sim N(X_i\beta, \Sigma)$:

$$Y_{ij} = \mathbb{1}(Y_{ij}^* > 0)$$

(2)

where $\beta$ is the environmental coefficient matrix and $\Sigma$ the covariance matrix. Then the probability to observe $Y_i$ is:

$$Pr(Y_i | X_i\beta, \Sigma) = \int_{A_{iJ}} \ldots \int_{A_{i1}} \phi_J(Y_i^*;\ X_i\beta, \Sigma) dY_{i1}^* \ldots dY_{iJ}^*$$

(3)

With the interval $A_{ij}$:

$$A_{ij} = \begin{cases} (-inf, 0] & Y_{ij} = 0 \\ [0, +inf] & Y_{ij} = 1 \end{cases}$$

(4)

The main computational issue of the full-MVP (Eq. 3, $\phi$ is the density function of the multivariate normal distribution) is that calculating the probability of $Y_i$ requires to integrate over $Y_i^*$, which has no closed analytical expression for more than two species ($J > 2$). This makes the evaluation of the likelihood computationally costly when J >>1, and motivates the search for an efficient numerical approximation of Eq. 3.

To see how this approximation can be achieved, note that Eq. 3 can be expressed more generally as:

$$\mathcal{L}(\beta, \Sigma; Y_i, X_i) = \int_\Omega \prod_{j=1}^{J} Pr(Y_{ij} | X_i\beta + \xi)\, Pr(\xi|\Sigma)\, d\xi$$

(3)

In sjSDM, we approximate the integral in Eq. 3 by *M* Monte-Carlo samples from the multivariate normal species-species covariance. Because each of these samples is drawn from the covariance, the covariance term in Eq. 3 is integrated out, and we can calculate the remaining part of the likelihood as in a univariate case, and use the average of the *M* samples to get an approximation of Eq. 3:

$$\mathcal{L}(\beta, \Sigma;\, Y_i, X_i) \approx \frac{1}{M} \sum_{m=1}^{M} \prod_{j=1}^{J} P(Y_{ij} | X_i \beta + \xi_m)$$

$$\xi_m \sim MVN(0, \Sigma)$$

(4)

This approximation of the MVP was first proposed by Chen *et al.* 2018 in the context of fitting deep neural networks with an MVP response structure. Its most notable computational advantage over other existing approximations to the MVP problem, such as the Geweke-Hajivassiliou-Keane (GHK) algorithm (Hajivassiliou & Ruud 1994), is that the Monte Carlo sampling in Eq. 4 can be parallelized. This is especially efficient when performing calculations on GPUs rather than CPUs, due to their much higher number of cores (the idea itself of using the GPU for expensive computational tasks is not new in ecology per se, see Golding 2019). The GHK algorithm, on the other hand, is based on a recursive and thus non-independent importance sampling procedure, which means that the sampling cannot be parallelized.

For sjSDM, we implemented this approximation, which was previously only used in the deep learning literature, to the standard generalized linear MVP, which means that we conform to the model structure typically used in this field and can profit from all benefits associated with parametric models. Note that we use here an approximation of the probit link which we found in iterim results more beneficial for stochastic gradient descent than the analytical probit link itself (see Supplementary Information for details).

We implemented the method in an R package (https://github.com/TheoreticalEcology/s-jSDM), using the Python package PyTorch, which was particularly designed for deep learning (Paszke *et al.* 2017), and the R package reticulate, which allows us to use Python packages from within R (Ushey *et al.* 2019) to run PyTorch from within R. This setup allows us to leverage various sophisticated numerical algorithms from PyTorch, including the possibility to switch between efficiently parallelized CPU and GPU calculations, efficient parallelization and the ability to obtain analytical gradients (via automatic derivatives) of the MVP likelihood with the latent covariance structure marginalized out via the Monte-Carlo ensemble. The combination of efficient parallelization and analytical derivatives of the Monte-Carlo approximated likelihood makes finding the maximum likelihood estimate (MLE) for the full-MVP model extremely fast, despite the large number of parameters to optimize.

Outsourcing the Monte-Carlo approach to a GPU solves the issue of computational speed (as we show below), but it does not yet solve the problem that the covariance matrix has a very large number of parameters, which raises the problem of overfitting when the method is used on small datasets. To address this, we penalized the actual covariances in the species-species covariance matrix, as well as the environmental predictors, with a combination of ridge and lasso penalty (elastic net, see Zou & Hastie 2005, more details below). Our R package includes a function to tune the strength of the penalty for each model component separately via cross-validation.

The here-tested implementation of sjSDM only considers binary (presence-absence) data, but there are several routes to extend the approach also to count and proportional data. To a large extent, those are already implemented in our R package. Currently supported are count (Poisson distribution with log-link), presence-absence (Binomial distribution with logit and probit links), and normal data (Multivariate normal distribution). An advantage that is often attributed to JSDMs based on the LVM is the model-based ordination, which is not possible with sjSDM since it is based on the originally proposed MVP model without latent environmental variables. However, new ordination techniques with a focus on co-occurrence patterns (e.g. Popovic *et al.* 2019) could complement sjSDM in practical analyses.

# Benchmarking our method against state-of-the-art JSDM implementations

To benchmark our approach, we used six datasets from Wilkinson *et al.* 2019, a recent JSDM benchmark study (Table S1). Covariates were centered and standardized. Since BayesComm was the fastest JSDM implementation in Wilkinson *et al.* 2019, we rerun BayesComm with the same parameters as in Wilkinson *et al.* 2019 on our hardware and compared it to sjSDM. Using this data has the advantage that we can also compare our results to Wilkinson *et al.* 2019.

Additionally, we simulated new data from an MVP (Eq. 2), varying the number of sites from 50 to 500 (50, 70, 100, 140, 180, 260, 320, 400, 500) and the number of species as a percentage (10%, 30% and 50%) of the sites (e.g. the scenario with 100 sites and 10% results in 10 species). In all simulations, the species' environmental preference was described for five environmental covariates (beta), which was randomly selected. For each scenario was sampled 5 times. Here, all species had species-species associations, i.e., the species-species covariance matrices were not sparse (for details, see Supporting Information S1).

To compare our model to existing JSDM software packages, we selected BayesComm (version 0.1-2, Golding & Harris 2015), the fastest non-latent MVP according to Wilkinson *et al.* (2019), and two state-of-the-art latent-variable JSDM implementations: Hmsc (version 3.0-4, Tikhonov *et al.* 2019b), which uses MCMC sampling, and gllvm (version 1.2.1, Niku *et al.* 2020), which uses variational Bayes and Laplace approximation to fit the model. We used the default parameter settings for all three methods which were in line with other recent JSDM benchmarks (details see Supporting Information S1).

To estimate the influence of the number of Monte-Carlo samples on the error of the MVP approximation, we used 100 Monte-Carlo samples for each species when run on the CPU and 1,000 Monte-Carlo samples for each species when run on the GPU for sjSDM. In the following, we will call sjSDM when run on the GPU, GPU-sjSDM, and when run on the CPU, CPU-sjSDM. Since GPUs might be not commonly available, we wanted also to estimate the efficiency of running sjSDM on the CPU.

To assess the predictive performance of the models, we calculated the average area under the curve (AUC) over all species and 5 independent replicates for each scenario of a hold-out dataset (same size as the dataset used for fitting the model). The AUC measures the capability of the model to distinguish between absence and presence of species. To calculate the accuracy of the estimated species associations and environmental coefficients, we used root mean squared error and the accuracy of the coefficients' signs, again averaged over all species and replicates.

## Regularization to infer sparse species-species associations

For the previous benchmark, we simulated data under the assumption that all species interact. While this assumption may or may not be realistic, it is generally desirable for a method to work well also when there is only a small number of associations, i.e. when the species-species covariance matrix is sparse. We were particularly interested in this question because we conjectured that the LVM approach imposes correlations on the species-species associations that makes it difficult for LVM models to fit arbitrarily sparse covariance structures.

We therefore simulated the same scenarios, but with 95% sparsity in the species-species associations. To adjust our model to such a sparse structure, we applied an elastic net shrinkage (Zou & Hastie 2005) on all off-diagonals of the covariance matrix. Following Zou & Hastie 2005, lambda (the regularization strength) and alpha (the weighting between LASSO and ridge) were tuned via 5-folded cross-validation in 40 random steps. In the MVP model, species are correlated within sites and not between sites. By using entire sites in the holdouts, we applied a blocked CV across the species as the non-independent species are kept together (Roberts *et al.* 2017). On the other hand, to account for correlations among sites (a realistic scenario in empirical datasets) which is equivalent to spatial auto-correlation, one could additionally block sites (e.g. by using the blockCV package, Valavi *et al.* 2019).

Here, we used 1,000 samples for the MVP approximation in sjSDM, because interim results showed that the approximation error can be crucial for the tuning. For BayesComm, and gllvm we used the default settings (see details and additional comments in Supporting Information S1). For Hmsc, also the default settings were used and following Tikhonov *et al.* 2020 associations with less than 95% posterior probability being positive or negative were set to zero.

To measure the accuracy of inferred species-species associations for this benchmark, we normalized the covariance matrices to correlation matrices and calculate the true skill statistic (TSS = Sensitivity + Specificity - 1, Allouche *et al.* 2006) by transforming the true and predicted associations into two classes: all absolute associations smaller than 0.01 were assigned to class '0' and all absolute associations greater than 0.01 were assigned to class '1'. That way, a two-class classification problem was obtained and the TSS was calculated.

# Case study – Inference of species-species associations from eDNA

To demonstrate the practicability of our approach, we fitted our model to an eDNA community dataset from a published study that sampled 130 sites across Denmark (for details on the study design see Brunbjerg *et al.* 2017; for data and bioinformatics see Frøslev *et al.* 2019). On each site, eight environmental variables were recorded: precipitation, soil pH, soil organic matter, soil carbon content, soil phosphorous content, and mean Ellenberg's indicator values (light condition, nutrient status, and soil moisture) based on the plant community. Frøslev *et al.* 2019 identified by eDNA sequencing (81 samples per site) 10,490 OTU. We followed Frøslev *et al.* 2019 and removed five sites with less than 4 OTCU presences. We used only OTU occurring at least three times over the remaining 125 sites, which reduced the overall number of OTU from 10,490 to 3,649 OTU. All eight environmental variables were used in our analysis as main effects on the linear scale. The final dataset consisted of 3,649 OTU co-occurrences over 125 sites with eight environmental variables.

For this analysis, we set the regularization for the z-transformed environmental predictors to lambda = 0.1, and alpha = 0.5 (equal weighting of ridge and LASSO regularization). The regularization for the covariances of the species-species associations were tuned over 40 random steps (independent samples from the hyper-parameter space) and with leave-one-out cross-validation. For each of the resulting 40*125 = 5,000 evaluations, we fitted a GPU-sjSDM model in 150 iterations (with a batch size of 12 and 125 site, one iteration consists of 100 optimization steps (see Bottou, 2010)), 3,649 *3,649 weights for the covariance matrix (see Supplementary Information S1 for details about the parametrization of the covariance matrix in sjSDM), with batch size of 8 and learning rate of 0.001 (the size of the update of the parameters in one optimization step).

# Results

## Method validation and benchmark against state-of-the-art JSDMs

### Computational speed

On a GPU, our approach (GPU-sjSDM) required under 3 seconds runtime for any of our simulated data with 50 to 500 sites and 5 to 250 species. When run on CPUs only (CPU-sjSDM), runtimes increased to a maximum of around 2 minutes (Fig. 1A, Fig. S1). In comparison, Hmsc had a runtime of around 7 minutes for our smallest scenario and increase in runtime exponentially when the number of species exceeded 40 (Fig. 1A). BayesComm was slightly faster than Hmsc but scaled worse than Hmsc to large data sizes (Fig. S1). Gllvm achieved low runtimes, equivalent and sometimes better than our method for small data (< 50 species), but beyond that runtime started to increase exponentially as well, leading to runtimes >10 min for our larger test cases (Fig. 1A).

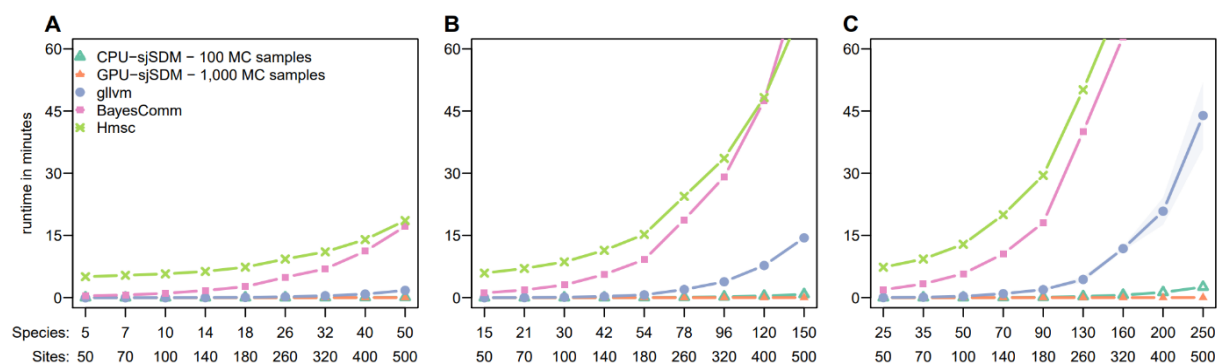

**Figure 1:** Runtime benchmarks for GPU-sjSDM, CPU-sjSDM, gllvm, BayesComm, and Hmsc fitted to simulated data with 50 to 500 sites (dense species-species association matrices) and the number of species set to A) 0.1, B) 0.3 and C) 0.5 times the number of sites. All values are averages from 5 simulated datasets. To estimate the inference error of the Monte-Carlo approximation, GPU-sjSDM was fitted with 1,000 and CPU-sjSDM with 100 MC-samples for each species.

Because of the runtime limitations of the other approaches, we calculated big data benchmarks only for GPU-sjSDM. The overall runtimes for GPU-sjSDM increased from under one minute for 5,000 sites to a maximum of around 4.5 minutes for 30,000 sites (Fig. 2). GPU-sjSDM showed greater runtime increases when increasing numbers of sites, while the numbers of species (300, 500, and 1,000 species in each scenario) had only small effects on runtimes (Fig. 2).

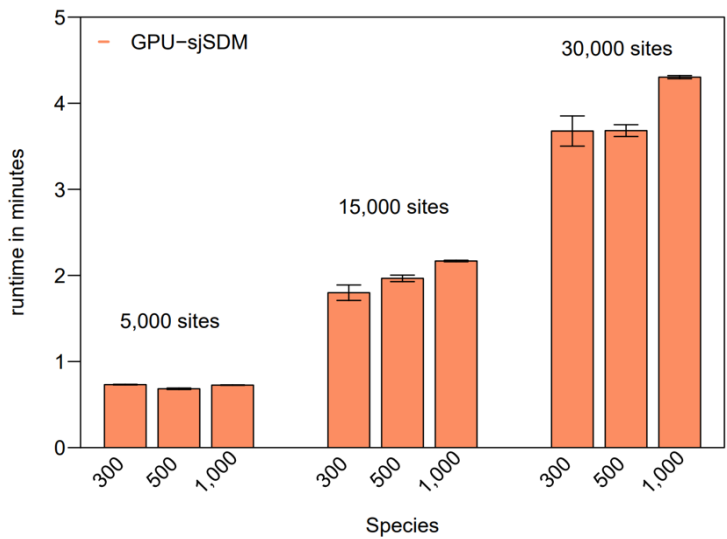

**Figure 2:** Benchmark results for sjSDM on big community data. We simulated communities with 5,000, 15,000, and 30,000 sites and for each set of 300, 500, and 1,000 species.

For the empirical benchmarking datasets from Wilkinson *et al.* (2019), the CPU based CPU-sjSDM model achieved a 3.8 times lower runtime for the bird dataset and 23 times lower runtime for the butterfly dataset than BayesComm, the fastest JSDM in Wilkinson *et al.* 2019. The GPU-sjSDM run on the GPU achieved a 500 times lower runtime for the bird dataset and a 150 times lower runtime for the butterfly dataset than did BayesComm, the previously fastest model in Wilkinson *et al.* 2019 (Table 1).

**Table 1:** Model runtimes in hours. Results for BayesComm against our new approach sjSDM (CPU and GPU version).

|  | Wilkinson *et al.* 2019 | | Our Approach | |
| --- | --- | --- | --- | --- |
| Dataset | Size (site * species) | BayesComm | CPU-sjSDM | GPU-sjSDM |
| Birds (Harris 2015) | 2,752 * 370 | 3.5 | **0.97** | **0.007** |
| Butterflies (Ovaskainen *et al.* 2016) | 2,609 * 55 | 0.15 | **0.01** | **0.001** |
| Eucalypts (Pollock *et al.* 2014) | 458 * 12 | <0.01 | **<0.001** | **<0.001** |
| Frogs (Pollock *et al.* 2014) | 104 * 9 | <0.002 | **<0.001** | **<0.001** |
| Fungi (Ovaskainen *et al.* 2010) | 800 * 11 | <0.02 | **<0.001** | **<0.001** |
| Mosquitos (Golding 2015) | 167 * 16 | <0.01 | **<0.001** | **<0.001** |

## Accuracy of the inference about species-environment and species-species associations

For simulated data with dense species-species association structures, BayesComm and sjSDM consistently achieved higher accuracy in the inferred species-species associations than the LVM models Hmsc and gllvm (Fig. 3 A-C). The accuracy of all methods decreased with an increasing proportion of species, to around 70% for the full-MVP models (sjSDM and BayesComm) and 60% for the LVM models (Fig. 3 A-C). Even for communities with 300 to 1,000 species, sjSDM achieved accuracies of 69% percent and higher (Table S4).

For environmental preferences (measured by RMSE), Hmsc showed slightly higher inferential performance when the number of sites was low (Fig. S4A, B) while all models performed approximately equal for a high number of sites (Fig. S4A, B).

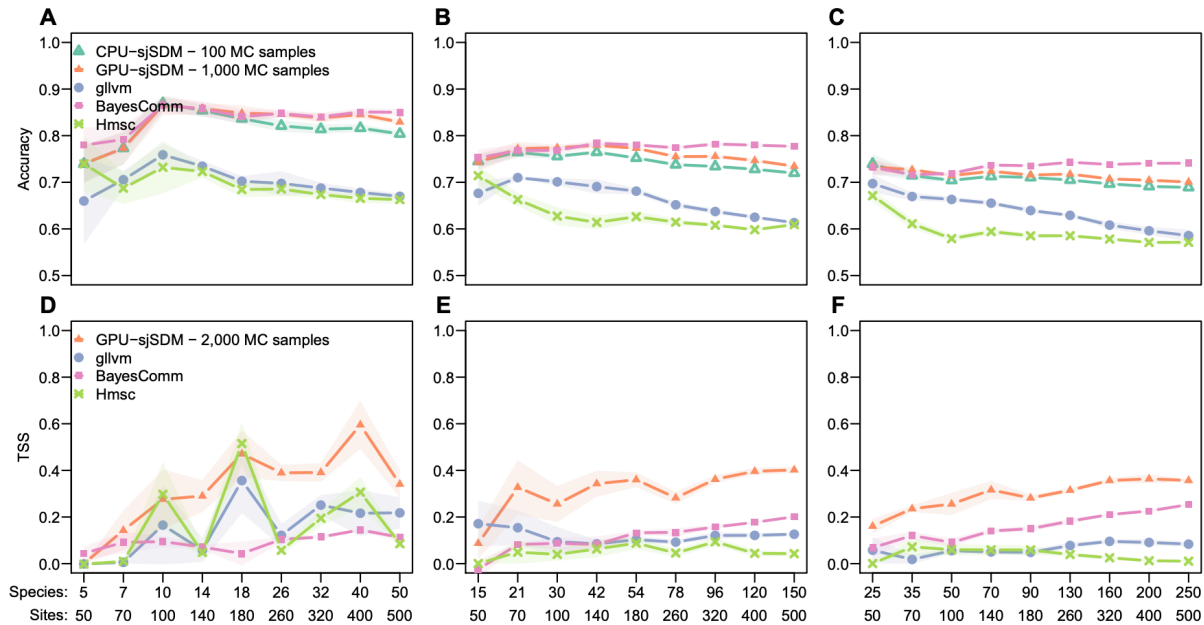

**Figure 3:** Inference performance of the inferred sparse and non-sparse species-species associations. Models were fitted to simulated data with 50 to 500 sites. All values are averages from 5 simulated datasets. A - C) The upper row shows the accuracies of matching signs (positive or negative covariance) for the estimated and true dense species-species association matrix. D - F) The lower row shows the accuracy of inferring non-zero species associations for sparse association matrices (95% sparsity), measured by the true skill statistic (absolute associations smaller than 0.01 were assigned the class '0' and absolute associations greater than 0.01 were assigned the class '1'). The number of species for were set to 0.1 (A, D), 0.3 (B, E) and 0.5 (C, F) times the number of sites. To estimate the inference error of the Monte-Carlo (MC) approximation, GPU-sjSDM was fitted with 1,000 and CPU-sjSDM with 100 MC-samples for each species.

For simulated data with sparse species-species association structures (95% sparsity), sjSDM achieved the highest TSS (up to 0.35-0.38 with 30% and 50% species, see Fig. 3 D-F). Hmsc showed for 10% species the second highest TSS (Fig. 3 D-F), but for 30% and 50% species together with gllvm the lowest TSS (a maximum of 0.1 TSS for 30% and 50% species). BayesComm showed in average the lowest TSS for 10% species, but for 30% and 50% species the second highest TSS (Fig. 3 D-F). The inferential performance regarding the environmental predictors showed the same pattern as for dense species-species associations. All models improved their environmental accuracy (Fig. S4C) and reduced RMSE as the number of sites increased (Fig. S4D).

## Predicting species occurrences

All models performed similarly in predicting species occurrences in the simulation scenarios, with predictive accuracies of around 0.75 AUC (Fig. 4).

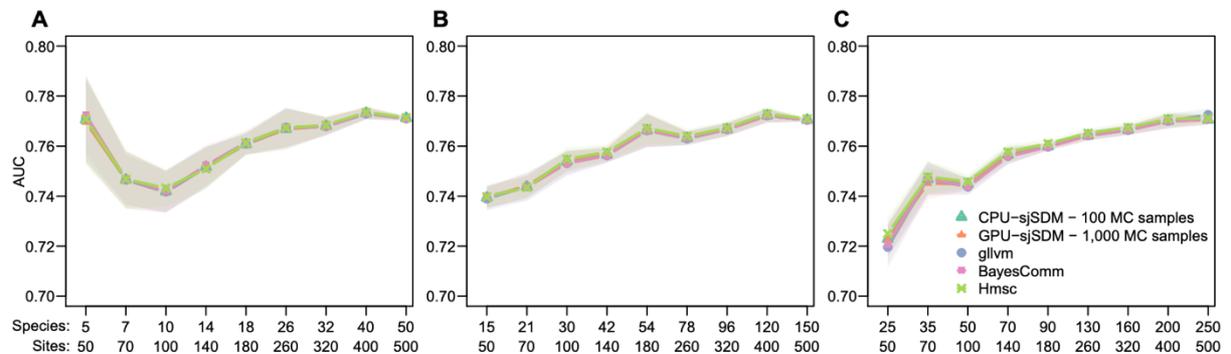

**Figure 4:** Predictive performance in simulated species distributions for GPU-sjSDM and CPU-sjSDM with gllvm, BayesComm, and Hmsc as references. Species distribution scenarios with A) 50 to 500 sites and 10%, B) 30% , and C) 50% species were simulated, on which the models were fitted (training). Models predicted species distributions for additional 50 to 500 sites (testing). Area under the curve (AUC) was used to evaluate predictive performance on holdout.

# Case Study - Inference of species-species associations from eDNA

In our eDNA case study with 3,649 OTU over 125 sites, we found that without regularization, sjSDM inferred the strongest negative OTU-OTU covariances among the most abundant species and the strongest positive OTU-OTU associations among the rarest OTU (Fig. 5a, b). When optimizing the regularization strength for the OTU-OTU associations via a leave-one-out cross-validation, positive and negative OTU-OTU associations changed somewhat, but the overall pattern stayed qualitatively constant (Fig. 5A, B). For the environmental covariates (a weak non-optimized regularization was used), we found that most OTU showed the highest dependency on ellenberg F (moisture), ellenberg L (light availability), and ellenberg N (nitrogen).

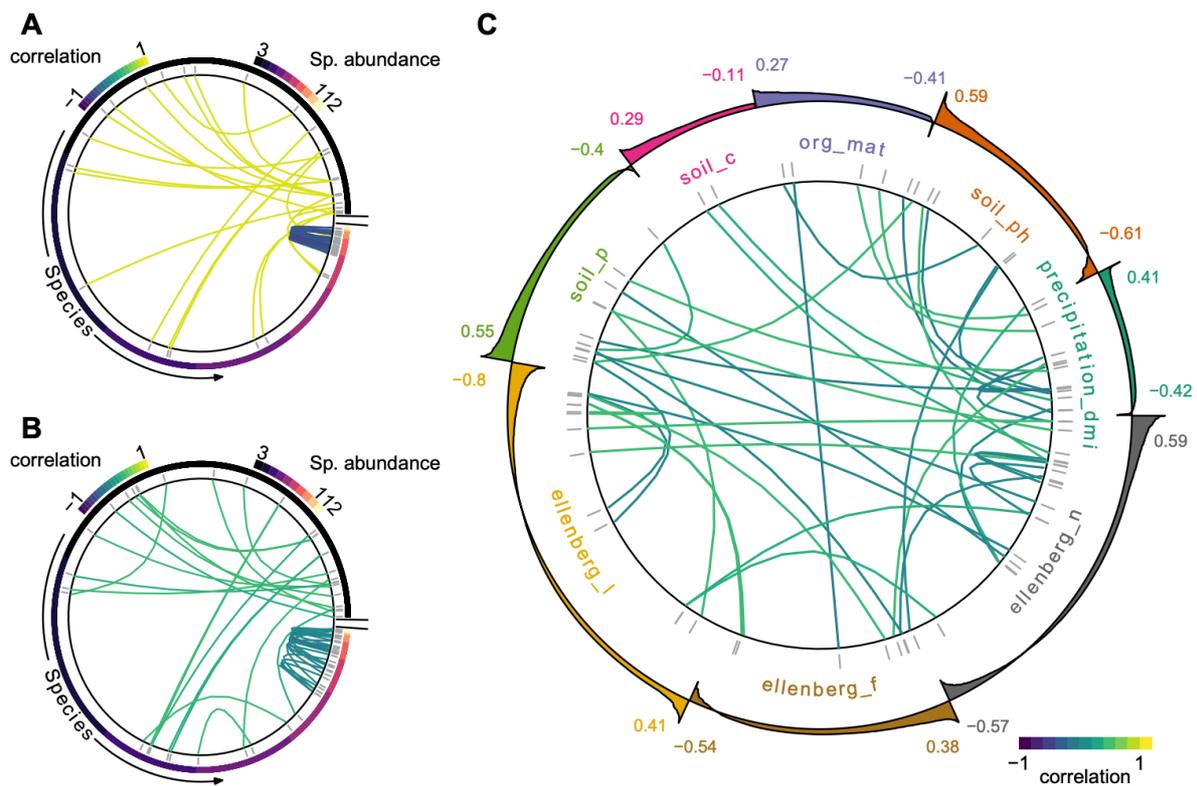

**Figure 5:** Inferred OTU associations and environmental preferences for the eDNA community data. The left column (panel A-C) shows OTU-OTU associations for A) no regularization and B) tuned regularization, with the 3,649 OTU sorted according to their summed abundance over 125 sites. The large panel C) shows the covariance structure of B), but with OTU sorted after their most important environmental coefficients (largest absolute environmental effect size; the outer ring shows the environmental effect distribution for the OTU within the group).

# Discussion

JSDM extend standard species distribution models by accounting also for species-species associations. Current JSDM software, however, exhibit computational limitations for the large community matrices, which limits their use for big community data that are created by novel methods such as eDNA studies and metabarcoding. Here, we presented sjSDM, a new numerical approach for fitting JSDMs that uses Monte-Carlo integration of the model likelihood, which allows moving calculations to GPUs. We show that this approach is orders of magnitude faster than existing methods (even when run on the CPU) and predicts as well as any of the other JSDM packages that we used as a benchmark. To avoid overfitting, especially when fitting sjSDM to hitherto computationally unrealistic eDNA datasets with thousands of species, we introduced a flexible elastic net regularization on species associations and environmental preferences. sjSDM inferred the signs of full association matrices and identified zero/non-zero entries in sparse species-species associations across a wide range of scenarios better than all tested alternatives. Advantages of BayesComm and sjSDM over LVM based JSDMs (Hmsc and gllvm) occurred in particular for sparse species-species associations, while improvement over latent-variable were visible for all tested species-species association structures.

## Computational performance

Whereas runtimes for Hmsc, BayesComm, and gllvm started to increase exponentially when the number of species exceeded around 100, sjSDM scaled close to linearly with the number of species regardless of whether we used GPU or CPU computations (Fig. 1a). A further advantage of sjSDM is that unlike in particular the MCMC algorithms used in BayesComm and Hmsc, it is highly parallelizable, which allows using efficiently the advantages of modern computer hardware such as GPUs. These two properties, scalability and parallelizability, make sjSDM the first and currently only JSDM software package that seems capable of analyzing big eDNA datasets (Humphreys *et al.* 2019; Tikhonov *et al.* 2019a; Wilkinson *et al.* 2019) on standard computers with acceptable runtimes.

We conceit that runtimes of the different JSDM implementations may depend on hyper-parameters such as the number of MCMC iterations in BayesComm or Hmsc, or the number of MC samples in sjSDM. Changing these parameters could affect results, however, increasing or decreasing MCMC iterations would only linearly shift the runtime curves (Fig. 1, S1). When we compare such a linear shift with the strong nonlinearity scaling of BayesComm and Hmsc, it seems unlikely that changes to the hyper-parameters could qualitatively change the results.

Moreover, sjSDM uses a Monte-Carlo approximation of the likelihood, but all other tested methods use approximations as well to obtain the inference. Neither our inferential results nor other indicators give us reasons to think that the approximation made by sjSDM is worse than that of competing algorithms. Specifically, increasing the number of Monte-Carlo samples for each species in sjSDM from 100 to 1,000 increased only slightly the inferential performance (Fig. 3A-C). Also, the excellent inferential accuracy of sjSDM across various tests does not suggest a large approximation error. We are therefore confident that our Monte-Carlo approximation is reliable.

State-of-the-art JSDM implementations offer a variety of extensions such as the inclusion of phylogeny, space, and traits (e.g. Hmsc, Tikhonov *et al.* 2020). Here, we demonstrate the advantage of sjSDM in estimating a simple MVP structure which is arguably the most basic and generic version of a JSDM and implemented by all packages, thus making it the logical common denominator to compare different packages. But sjSDM can be extended to include other structures that have been proposed in the literature, such as traits, as is done in gllvm by using the model-based fourth corner model (Brown *et al.* 2014; Niku *et al.* 2019, not supported in sjSDM yet). We also compared here only presence-absence occurrences but the sjSDM package supports normal, Poisson, and binomial responses, and has an option to model spatial coordinates. Options to account for space via spatial eigenvectors or a conditional autoregressive model (CAR) are currently explored. A question for all these points is of course if they interact beneficially with our MLE approximation, i.e., if we can optimize the MLE without having to resort to other integration methods (such as MCMC or Laplace approximations) for the added structures, which would negate the speed advantage of sjSDM.

## Inferential performance

All JSDM implementations showed similar performance in correctly inferring environmental responses, but the non-latent approaches sjSDM and BayesComm achieved significant higher accuracy in inferring the correct signs of species-species associations (Fig. 3A-C) and identifying sparse structures (Fig. 3D-F). It should be noted here that we tuned the regularization of sjSDM to improve the performance for sparse associations and the other JSDM might also benefit from tuning the regularization. BayesComm and Hmsc allow more restrictive priors to be specified on the covariance matrix (BayesComm) or on the factors loadings (Hmsc). However, the long runtimes of these JSDM implementations place time constraints on testing different prior specification. Moreover, BayesComm already achieved high TSS for sparse associations with default specifications, indicating superiority of highly parametrized JSDM over LVM for sparse structures (Fig. 3D-F).

The original idea of LVM algorithms was to the reduce the overall model complexity and the accompanying runtime, but apparently the constraints imposed by this structure create some bias that showed in particular for dense species association structures (compare Fig. 2A, Fig. S2). This is not particularly surprising, as similar phenomena have been found also for other approaches to covariance regularizations, for example in spatial models (Stein 2014).

It is difficult to estimate how important these biases are in practical applications, because we still know too little about the typical structure of species associations in real ecological data (Ovaskainen *et al.* 2017a). One might expect that associations in data generated by high-throughput technologies, which detect species already at very low densities, would be relatively sparse, or consist of a mix between sparse and non-sparse blocks for rare and common species (cf. Calatayud *et al.* 2019). Moreover, one would expect that LVM models would be particularly efficient if species associations follow the structure implemented in the LVM models. To test this, we also simulated data from an LVM structure, and fitted this data with sjSDM and the two LVM models (gllvm and Hmsc). Our results show that the LVM models indeed perform better than for such data than for our previously used general covariance matrices, but not better than sjSDM (Fig. S8, S8, S9).

A slight disadvantage of sjSDM is that it is more complicated to obtain parameter uncertainties, compared to JSDM implementations based on MCMC sampling such as BayesComm and Hmsc. The R implementation of sjSDM calculates Wald confidence intervals for all environmental predictors using PyTorch's automatic differentiation feature. However, we have currently no analytical option to calculate confidence intervals for the species-species associations. If these are needed, we propose using bootstrap samples.

## Implications and outlook for ecological data analysis

The JSDM structure has the potential to become the new default statistical approach for species and community observations that originate from eDNA and similar big community data. However, to fulfill this promise, we need statistical algorithms that computationally scale to big datasets, and deliver accurate inference, in particular for a large number of species or operational taxonomic units. Our results show that a combination of a scalable and parallelizable Monte-Carlo approximation of the likelihood, together with a shrinkage regularization of the species-species covariance, might be a configuration that can achieve both goals.

We see further need for research, particularly on the question of how to impose regularization, or other structures on the covariance. In principle, all software packages that we compared could include additional regularization methods, such as the elastic net employed in our approach. Better understanding the use of such statistical approaches is one promising route for further research. Another option would be to impose ecologically motivated structures on the species-species covariance matrix (e.g. Clark *et al.* 2017; Taylor-Rodríguez *et al.* 2017; Bystrova *et al.* 2021).

Another interesting question is how ecologist will use JSDMs, once they scale to big data. Many recent studies have stressed that JSDMs may improve predictions (e.g. Norberg *et al.* 2019), and indeed, from ecological theory, one would expect that species associations are important for accurate species occurrence predictions (Dormann *et al.* 2012; Wisz *et al.* 2013; Norberg *et al.* 2019). Despite different accuracy in inferring true species interactions structures (Fig. 3), we found similar predictive performances (Fig. 4) for JSDMs in our results. However, metrics such as AUC capture only the marginal predictive performance and not the joint predictive performance of JSDMs (Wilkinson *et al.* 2021).

This, however, raises the question whether reported increases in predictive performance of JSDMs are really due to their exploitation of a stable association structure, or simply arise from the higher model complexity of JSDMs, which allows fitting the data more flexibly. Further systematic benchmarks, where model structures on biotic and abiotic predictions are flexibly adopted (e.g. via machine learning approaches such as in Chen et al. 2018), and where indicators of joint predictive performance (Wilkinson *et al.* 2021) are used that are sensitive to covariances could help to better understand how important biotic interactions truly are for achieving high predictive performance.

A completely new type of information that JSDMs deliver to ecologists are species-species covariance estimates (Leibold *et al.* 2020). These could be used, for example, to test if the strength or structure of species associations varies with space or environmental predictors; or if spatial species associations correlate with local trophic or competitive interactions or traits (see generally Poisot *et al.* 2015). For regional studies, there is the prospect of extending the traditional variation partitioning (environment and space) (Cottenie 2005) to include biotic associations by using JSDMs (Leibold *et al.* 2020). Our results regarding the moderate, but significantly better than random accuracy of inferred covariance structures, even on datasets with hundreds of species, are encouraging for such a research program.

Recently, concerns about the usefulness of JSDM emerged. For instance, it is has been criticized that the species-species associations inferred by JSDM cannot always be linked to ecological interactions because of their symmetric nature (e.g. Zurell *et al.* 2018; Blanchet *et al.* 2020), that the associations may absorb missing environmental covariates (Poggiato *et al.* 2021) or that JSDM associations can be scale-dependent (see König *et al.* 2021; although so can ecological interactions, see Poisot *et al.* 2015). We acknowledge these properties but do not share the concerns. JSDM estimate associations between species after accounting for the environment. Such associations are not necessarily causal or mechanistic, and they are naturally influenced by unmeasured predictors, scale and other factors, but they are also influence by real species associations (e.g. Leibold *et al.* 2020, Fig. S10), and thus provide at least some useful ecological information beyond pure niche models. If more high-resolution dynamic data was available, we could use more precise (causal) methods to infer the direction of interactions (Barraquand *et al.* 2019; Momal *et al.* 2020), which likely match much closer to actual species interactions. Yet, for the static community data that makes up the bulk of the data available to ecologists today, these methods are not applicable, but JSDMs are and can provide additional information compared to existing alternatives.

## Conclusion

We presented sjSDM, a new method to fit JSDMs, and benchmarked it against state-of-the-art JSDM software. sjSDM is orders of magnitudes faster than current alternatives, and it can be flexibly regularized, which leads to overall superior performance in inferring the correct species association structure. We emphasize that the superior scaling holds also when using CPU computations, and that the possibility to move calculations on a GPU is only a further advantage of the algorithm. We provide our tool in an R package

(https://github.com/TheoreticalEcology/s-jSDM, available for Linux, MacOS, and Windows), with a simple and intuitive interface and the ability to switch easily between linear and nonlinear modeling, as well as between CPU and GPU computing. The R package also includes extensions for considering abundance data as well as spatial coordinates, and to partition the importance of space, environment, and species associations for predicting the observed community composition.

# Acknowledgments

We would like to thank Douglas Yu and Yuanheng Li for valuable comments and suggestions.

# Authors' Contributions

FH and MP jointly conceived and designed the study. MP implemented the sjSDM software, ran the experiments, and analyzed the data. Both authors contributed equally to discussing and interpreting the results, and to the preparation of the manuscript.

# Data Accessibility

The compiled datasets for runtime benchmarking (case study 1) are available as supporting information for Wilkinson *et al.* 2019. The eDNA dataset is available at https://github.com/tobiasgf/man_vs_machine. The analysis and the R package sjSDM is available at https://github.com/TheoreticalEcology/s-jSDM.

# Support information S1

## Code and data availability

The compiled datasets for runtime benchmarking (case study 1) are available as supporting information for Wilkinson *et al.* 2019. The eDNA dataset is available at https://github.com/tobiasgf/man_vs_machine. The analysis and the R package sjSDM is available at https://github.com/TheoreticalEcology/s-jSDM.

## Simulation scenarios

The MVP can be interpreted as individual GLMs connected by correlated residuals, which are sampled from a multivariate Gaussian, and with a probit link. Sites are notated with $i = 1, \ldots, M$; species with $j = 1, \ldots, K$; and environmental covariates with $n = 1, \ldots, N$. Environmental covariates and species responses (beta) were uniformly sampled (Eq. 1). The lower triangular covariance matrix was uniformly sampled (Eq. 2), the diagonal was set to one (Eq. 3) and multiplicated by the transposed lower triangular to get a symmetric positive definite matrix (Eq. 4). Afterwards, the covariance matrix was normalized to the range of [-1,1] (Eq. 6).

$$\beta, X \sim U(-1, +1) \tag{1}$$

$$\Sigma^{lower}_{j \neq j} \sim U(-1, +1) \tag{2}$$

$$\Sigma^{lower}_{j = j} = 1 \tag{3}$$

$$\Sigma = \Sigma^{lower} * \left(\Sigma^{lower}\right)^T \tag{4}$$

$$D = \sqrt{(diag(\Sigma))} \tag{5}$$

$$\Sigma' = diag(D)^{-1} \Sigma \, diag(D)^{-1}$$

$$Z_{ij} = \beta_{0j} + \sum_{n=1}^{N} X_{in} * \beta_{nj} + e_{ij} \tag{6}$$

$$e_i \sim MVN(0, \Sigma') \tag{7}$$

$$Y_{ij} = 1\,(Z_{ij} > 0) \tag{8}$$

(9)

Species responses consist of a linear species - environmental response and correlated residuals (Eq. 7). Following a probit link, responses higher than zero were set to one and the remaining to 0 (Eq. 9).

## Approximation of multivariate probit model

The multivariate normal PDF is given by $\phi$. The probability to observe $Y_i$ for $X_i$ by the environmental coefficient matrix $\beta$ and the covariance matrix $\Sigma$ is given by:

$$P(Y_i|X_i\beta, \Sigma) = \int_{A_J} \ldots \int_{A_1} \phi(Y^*, X_i\beta, \Sigma)\, dY_1^* \ldots dY_J^*$$

$$A_j = \begin{cases} (-inf, 0] & y_{ij} = 0 \\ [0, +inf] & y_{ij} = 1 \end{cases}$$

(10)

We can rewrite this as the cumulative density function ($\Phi$) of the multivariate normal distribution:

$$D_i = diag(2Y_i - 1)$$

$$\boldsymbol{\mu}_i = D_i(X_i\beta)$$

$$\Sigma^* = D_i \Sigma D_i$$

$$P(Y_i|X_i\beta, \Sigma) = \Phi(0|-\boldsymbol{\mu}_i, \Sigma^*)$$

(11)

To approximate the likelihood in sjSDM, we use a Monte-Carlo approach that was suggested in a slightly different context by Chen *et al.* 2018. In the following, we shortly sketch the idea.

With $r \sim N(0, \Sigma^*)$, we can rewrite Eq. 11 as:

$$\Phi(0| -\boldsymbol{\mu}_i, \Sigma^*) = Pr(r_i - \boldsymbol{\mu}_i \leq 0)$$

(12)

We can now reparametrize $\Sigma^* = V + \Sigma^r$ with $V$ as a diagonal, which means that the random variable $r$ depends on two other random variables $z \sim N(0, V)$ and $w \sim N(0, \Sigma^r)$ and following $r = z + w$ Eq. 12 is equal to:

$$\Phi(0|-\boldsymbol{\mu}_i^*, \Sigma^*) = Pr(\boldsymbol{r}_i - \boldsymbol{\mu}_i \leq 0)$$

$$= Pr(\boldsymbol{z}_i - \boldsymbol{w}_i \leq \boldsymbol{\mu}_i)$$

(13)

In that way we transform the individual Monte-Carlo samples with covariance matrix, and we can treat them as univariate samples and use the univariate normal CDF:

$$= \mathbb{E}_{w \sim N(0, \Sigma^r)}[Pr(z \leq (\boldsymbol{w}_i + \boldsymbol{\mu}_i)|\boldsymbol{w}_i)]$$

$$= \mathbb{E}_{w \sim N(\boldsymbol{\mu}_i, \Sigma^r)}\left[\prod_{j=1}^{J}\left(\frac{\Phi(w_{ij})}{\sqrt{V_{jj}}}\right)\right]$$

(14)

$V$ is used to rescale the univariate samples and w.l.o.g. we can set $V$ to identity matrix $I$ and estimate only $\Sigma^r$. Following this, the final approximation is:

$$\approx \frac{1}{M}\sum_{m=1}^{M}\prod_{j=1}^{J}\Phi(w_j^m)$$

(15)

We can rewrite Eq. 15 as, with $\Sigma^{1/2}$ being the square-root matrix of $\Sigma^r$:

$$\boldsymbol{w}_i^{(m)} = \boldsymbol{X}_i\beta + \Sigma^{1/2}\boldsymbol{z}_i^{(m)}, \boldsymbol{z} \sim N(0, I)$$

$$P(\boldsymbol{Y}_i|\boldsymbol{X}_i\beta, \Sigma) \approx \frac{1}{M}\sum_{m=1}^{M}\prod_{j=1}^{J}\Phi\left(w_{ij}^{(m)}\right)Y_{ij} + (1 - Y_{ij})\left(1 - \Phi\left(w_{ij}^{(m)}\right)\right)$$

(16)

For optimizing the parameters, we then use the automatic derivates implemented in PyTorch to find the gradient for each Monte-Carlo particle, and average gradients of all particles to obtain a gradient for the optimizer. In short, the core of the algorithm is to generate an approximation of the gradient of the likelihood by drawing from the multivariate normal distribution in the MVP model, and propagating the calculations for the resulting draws through the entire model structure.

We said earlier that the random variable $r$ consists of actual two random variables $z$ and $w$, and thus the covariance matrix $\Sigma^*$ can be decomposed into $\Sigma^* = I + \Sigma^r$. For the actual sampling from the covariances ($\Sigma^r$), we can use the square root matrix:

$$\Sigma^* = I + \Sigma^{1/2} * \left(\Sigma^{1/2}\right)^T$$

(17)

The square root matrix $\Sigma^{1/2}$ has dimensions $J$ (number of species) $* d$ and is the actual parameter matrix we optimize in sjSDM. $d \ll J$ corresponds to a low-rank parametrization, increasing $d$ increases the overall number of parameters parametrizing $\Sigma^r$. The advantage of this re-parametrization is that $\Sigma^{1/2}$ needs not to be symmetric and positive definite, which means we can use it directly for sampling without, for example, using a Cholesky decomposition which is computationally expensive and numerically unstable for high $J$. Moreover, $d \ll J$ can be seen as another way to regularize the covariance matrix (it is similar to the LVM). However, in sjSDM we use an elastic-net regularization on the individual entries of $\Sigma^*$ (we could also use the precision matrix of $\Sigma^*$ here (which is implemented in the sjSDM package)) and we found in interim results that for $d \ll J$ the elastic-net regularization does not work properly (i.e. all entries are regularized to 0 or not). We assume that for $d \ll J$ the covariances in $\Sigma^*$ are not 'independently enough' parametrized (not in the statistical sense) since the regularization of individual covariances of $\Sigma^*$ leads to an indirect regularization of $\Sigma^{1/2}$. To use elastic-net regularization we set $d = J/2$ as default in the sjSDM implementation, but future research may be needed to explore the trade-off between regularization via $d$ and regularization via elastic-net on the covariance matrix.

The model itself is optimized via stochastic gradient descent (Bottou 2010) which means that the estimates of the model are always updated only on a random batch of the dataset and thus one iteration (called epoch in the deep learning field) consists of the number of optimization steps necessary to go once through the full dataset (e.g. for 100 observations and a batch size of 10, an iteration would consists of 10 optimization steps).

The univariate probit link ($\Phi$ in Eq. 16) can be approximated by scaling the logit link: $\Phi(x) \approx F(x * 1.7)$ (Baker & Kim 2004; Savalei 2006), which we found in interims results more beneficial than the analytic probit link. Stochastic gradient descent is numerically often unstable, so it is not uncommon in machine learning approaches to adjust activation functions (Xu *et al.* 2015; Zhao *et al.* 2017) and model structures to improve convergence properties of the algorithm and we assume that the approximation of the probit via scaling of the logit link is numerically more stable than the actual analytic probit link.

# Runtime on case study

We used compiled datasets from Wilkinson *et al.* 2019 (Table S1).

**Table S1:** Compiled datasets that were taken from Wilkinson *et al.* 2019

| Dataset | Original paper | Species | Sites | Covariates |
| --- | --- | --- | --- | --- |
| Birds | Harris 2015 | 370 | 2,752 | 8 |
| Butterflies | Ovaskainen *et al.* 2016 | 55 | 2,609 | 4 |
| Eucalypts | Pollock *et al.* 2014 | 12 | 458 | 7 |
| Frogs | Pollock *et al.* 2014 | 9 | 104 | 3 |
| Fungi | Ovaskainen *et al.* 2010 | 11 | 800 | 15 |

| Mosquitos | Golding 2015 | 16 | 167 | 13 |

# Model settings and computational environment for the benchmarks

This section provides a more detailed explanation about model settings and the computer setup under which our benchmarks were performed (See Table S1 for an overview). Unless stated otherwise, we used default settings for parameters.

**Table S2:** Overview of the used approaches

| Model | Optimization type | Package |
|---|---|---|
| Multivariate probit model | MLE | sjSDM |
| | MCMC | BayesComm |
| Latent-variable model | MCMC | Hmsc |
| | Variational bayes / Laplace approximation | gllvm |

## BayesComm

*BayesComm* models were fitted with 50,000 MCMC sampling iterations, with two chains, thinning = 50, and burn-in of 5000. Prior were not changed from default: normal prior on regression coefficients b ~ N(0;10) and an inverse Wishart prior on the covariance matrix.

## Hmsc

Hmsc models were fitted with 50,000 MCMC sampling iterations, with two chains, thinning = 50, and burn-in of 5000. Since the two chains were not run in parallel (although it is supported by Hmsc), the measured runtime was halved. The number of latent variables in Hmsc are automatically inferred by gamma shrinkage prior. The shrinkage priors of Hmsc were not changed from default.

We note that there is the option to tune regularization via shrinkage priors in Hmsc: $a_1$ regularizes the lower triangular of the species association and $a_2$ regularizes the number of latent variables (see Bhattacharya & Dunson 2011). We acknowledge that this might improve accuracy of Hmsc inference. On the other hand, it should be noted that a) these settings were not tuned in recent benchmarks and are likely not tuned by users either. b) the runtime of tuning several combinations would be not practicable (see our results) and c) it is to be expected that a low $a_2$ results in a higher accuracy but then the LVM approach would approximate the MVP model and that would contradict the LVM's unique characteristic.

## gllvm

gllvm models were fitted as binomial models with probit link. The number of latent variables were increased from 2 to 6 with the number of species. If default starting values = "res" caused an error, model was re-run with starting values = "zero" and if another error occurred, the model was re-run with starting values = "random". Run time was measured individually, not as a sum over possible model fitting tries.

## sjSDM

sjSDM models were fitted with 50 iterations (epochs) and a batch size of 10% number of sites. Learning rate was set to 0.01. 50% number of species were set for *d* for the parametrization ($J * d$ weights) of the covariance matrix (see section about the approximation, default in the sjSDM R-package). For sparse species association matrices, 50 iterations with a learning rate of 0.03 were used and the regularization of the species-species covariances were tuned in 40 random steps and 5-folded cross-validation. 2,000 Monte-Carlo samples were used for the MVP approximation because interim results showed that with too less samples the variance of the models' log-Likelihoods might interfere with the comparison.

**Table S3:** Model settings of JSDM implementations Hmsc, BayesComm, gllvm, and sjSDM.

| Model | Parameter | Value |
|---|---|---|
| Hmsc | Iterations (MCMC samples) | 50,000 |
| | Burnin | 5,000 |
| | Thinning | 50 |
| | Number of chains | 2 |
| | Distribution | Binomial with probit link |
| BayesComm | Iterations (MCMC samples) | 50,000 |
| | Burnin | 5,000 |
| | Thinning | 50 |
| | Number of chains | 2 |
| | Distribution | Binomial with multivariate probit link |
| gllvm | Number of latent variables | 2 for 10% species / site proportion |
| | | 3 for 30% species / site proportion |
| | | 4 for 50% species / site proportion |
| | Distribution | Binomial with probit link |
| | Starting values | 'res' (residulas). If model did not converge, retry with 'zero' and if model still didn't converge, retry with 'random' |
| sjSDM (non-sparse association matrices) | Learning rate | 0.1 |
| | Iterations | 50 |
| | MC-samples for each species | 100 (CPU); 1,000 (GPU) |
| | Distribution | Binomial with multivariate logit link |
| sjSDM (sparse associations) | Learning rate | 0.01 |
| | Iterations | 50 |
| | MC-samples for each species | 2,000 (GPU) |
| | Random tuning steps | 50 |
| | n-folded cross-validation | 5 |
| | Lambda range (regularization strength) | [9.765e-5, 0.063] |
| | Alpha (weighting between LASSO and ridge) | [0.0, 1.0] |

## Computer setup

All the computations were performed on the same workstation (two Intel Xeon Gold 6128 CPU @3.40 GHz) and the number of cores and threads were restricted to 6. GPU computations were carried out on a NVIDIA RTX 2080 Ti. All CPU models had access to 192 GB RAM and the GPU models to 11 GB GPU RAM). Analyses were conducted with the statistical software R and Python (Python Software Foundation. Python Language Reference, version 3.8.1. Available at http://www.python.org)

## Additional results

### Run time scaling of the algorithms in log plots

In the main paper, we provide the benchmarks on a linear scale. Below, we also provide then in log format, which demonstrates that many other software packages, including the CPU version of sjSDM scale exponentially, while G-sjSDM scales sub-exponentially for the scenarios that we tested (Fig S1).

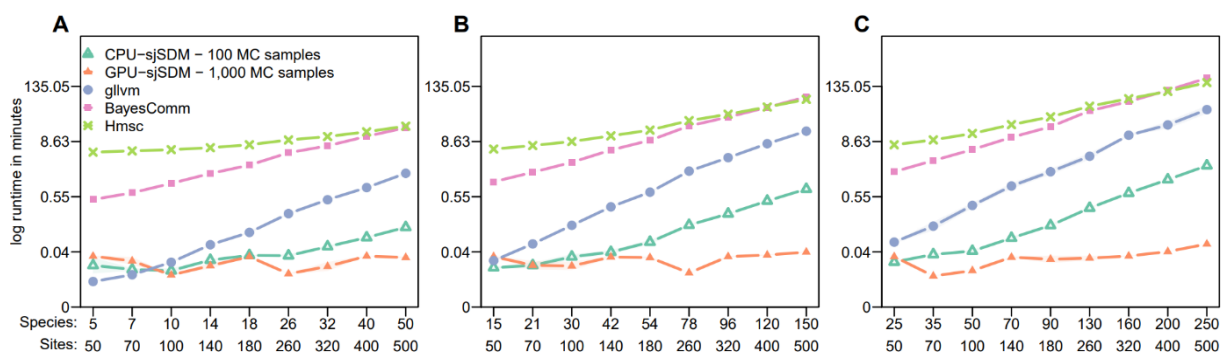

**Figure S1:** Results for computational log runtime benchmarking of G-sjSDM, C-sjSDM, gllvm, BayesComm, and Hmsc jSDM implementations. Models were fitted to different simulated SDM scenarios: 50 to 500 sites with A) 10%

, B) 30%, and C) 50% number of species (e.g. for 100 sites and 10% we get 10 species). For each scenario, ten simulations were sampled, and results were averaged. Due to high runtimes, runs for BayesComm, gllvm and Hmsc were aborted at specific points.

## Additional results

We additionally calculated the root mean squared error (RMSE) between the true simulated and estimated association matrices (Fig. S2). The pattern we found is consistent with the accuracy of the signs of the covariances. For dense and sparse association matrices, BayesComm and sjSDM achieved the lowest RMSE in all scenarios (Fig. S2).

Moreover, we calculated the true skill statistic (TSS) for different thresholds (0.005, 0.01, 0.05, and 0.1, entries below the threshold are classified as 0) when evaluating the accuracy in inferring zero and non-zero entries in the covariance matrices (Fig. S3). We found that sjSDM and BayesComm achieved the highest TSS for thresholds up to up to 0.05. With a threshold of 0.05, gllvm achieved a similar TSS as sjSDM and BayesComm. It seems, that the performance of BayesComm and especially for sjSDM decreases for higher thresholds, however, with increasing thresholds higher non-zero values are getting classified as zero resulting in non-sparse association matrices, thereby the decrease in TSS.

## Additional results of sjSDM on large scale datasets

Beside the runtimes for the large-scale datasets (see main paper), we also calculated the accuracy of the matching signs of predicted and true parameters for the association matrix and environmental coefficients (Table S4). Moreover, the RMSE for the environmental coefficients were calculated (Table S4).

Overall, we found that the association accuracy decreased from 300 to 1000 species with the different number of sites, but overall, the association accuracy increased with the number of sites (Table S4). The accuracy of environmental coefficients was close to 1.0 in all scenarios and the RMSE for the environmental coefficients was close to zero in all scenarios (Table S4).

**Table S4:** Accuracy of matching signs for estimated associations, environmental coefficients, and root mean squared error (RMSE) for environmental coefficients

| Sites | 5,000 | | | 15,000 | | | 30,000 | | |
|---|---|---|---|---|---|---|---|---|---|
| Species | 300 | 500 | 1000 | 300 | 500 | 1000 | 300 | 500 | 1000 |
| Covariance accuracy | 0.744 | 0.721 | 0.688 | 0.750 | 0.727 | 0.693 | 0.750 | 0.728 | 0.692 |
| Env accuracy | 0.988 | 0.985 | 0.987 | 0.990 | 0.989 | 0.991 | 0.990 | 0.990 | 0.990 |
| Env RMSE | 0.040 | 0.037 | 0.035 | 0.034 | 0.029 | 0.026 | 0.033 | 0.029 | 0.025 |

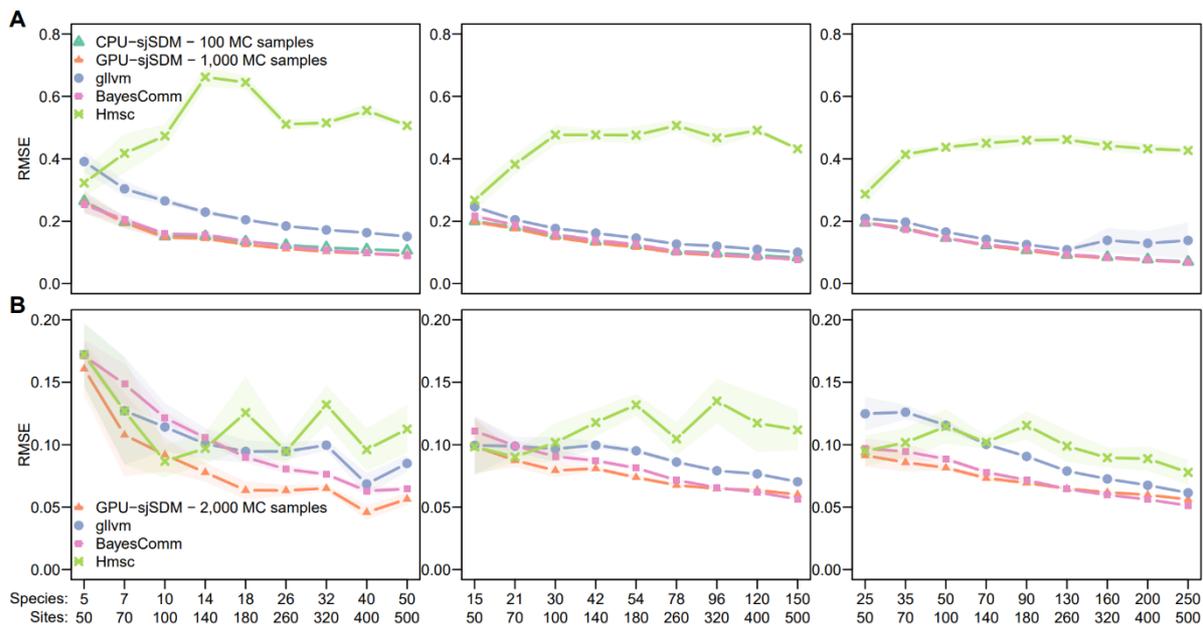

**Figure S2:** Root mean squared error of the inferred A) non-sparse and B) sparse species-species associations. Models were fitted to simulated data with 50 to 500 sites and the number of species set to 0.1, 0.3 and 0.5 times the number of sites. All values are averages from 5 simulated datasets.

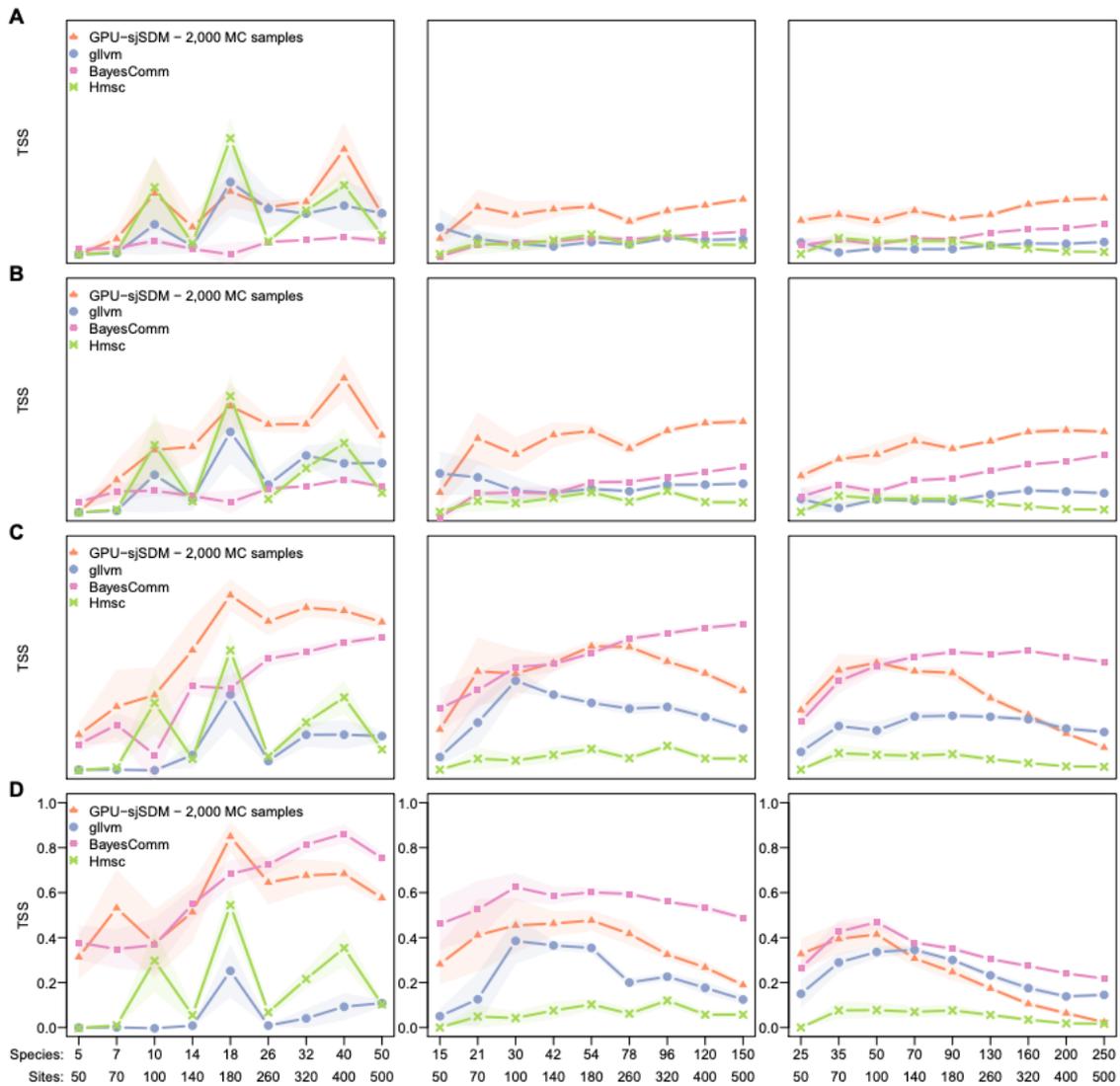

**Figure S3:** True skill statistic (TSS) for sparse absolute species-species associations with different thresholds: A) 0.005, B) 0.01, C) 0.05, and D) 0.1. Models were fitted to simulated data with 50 to 500 sites and the number of species set to 0.1, 0.3 and 0.5 times the number of sites. All values are averages from 5 simulated datasets.

## Accuracy of inferring environmental parameters

For non-sparse species-species associations, all models inferred with high accuracy true signs of the environmental predictors and achieved similar RMSE errors estimated environmental parameters (Fig. S4 A-B).

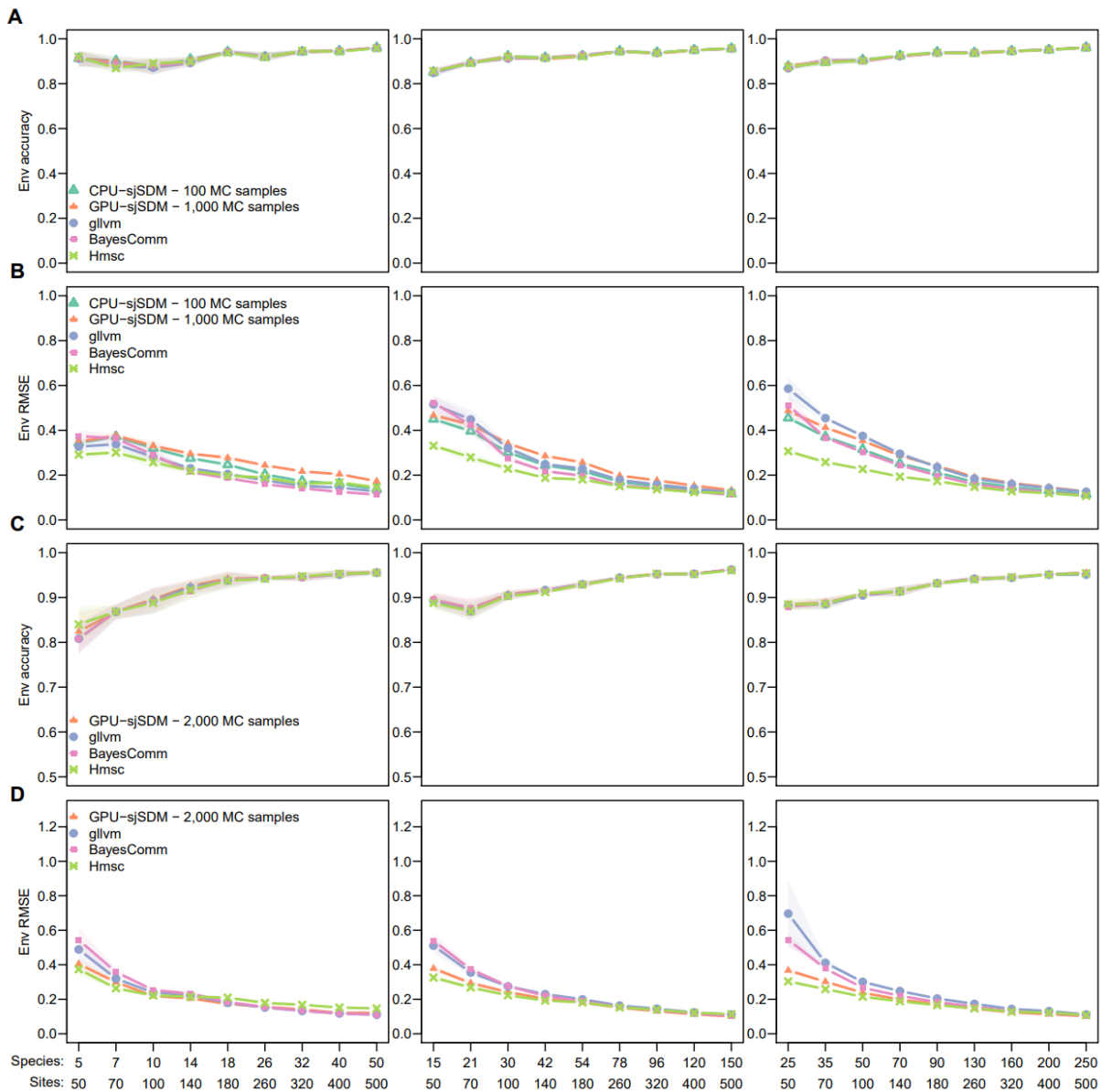

**Figure S4:** Results for inferential benchmarking of G-sjSDM with gllvm, BayesComm, and Hmsc as references. Models were fitted to different simulated SDM scenarios: 50 to 500 sites with 10% (first column), 30% (second column) and 50% (third column) number of species to site proportions (e.g. for 100 sites and 10% we get 10 species). For each scenario, 5 simulations were sampled, and results were averaged. A) and B) show the environmental coefficient accuracy (matching signs) and the corresponding RMSE with full species-species association matrices. C) and D) show the environmental coefficient accuracy (matching signs) and the corresponding RMSE with sparse (50% sparsity) species-species association matrices.

For sparse species associations, the models achieved similar performances in inferring the environmental parameters (Fig. S4 C-D).

## Convergence check

To check convergence of sigma and the betas, the potential scale reduction factors (psrf) for Hmsc and BayesComm in the simulation scenarios were calculated (two chains, burn-in = 5000 and 50,000 sampling iterations). We found no psrf > 1.2 for BayesComm, but for Hmsc in most simulation scenarios at least for one parameter (beta or lambda (factor loadings)) a psrf > 1.2 (Fig. S5).

In case of Hmsc, we removed the psrf factors for the last latent factor loading because interim results showed that the psrfs for the last factor loading were on average always higher than 1.2, although they were estimated to be very small (perhaps a numerical issue).

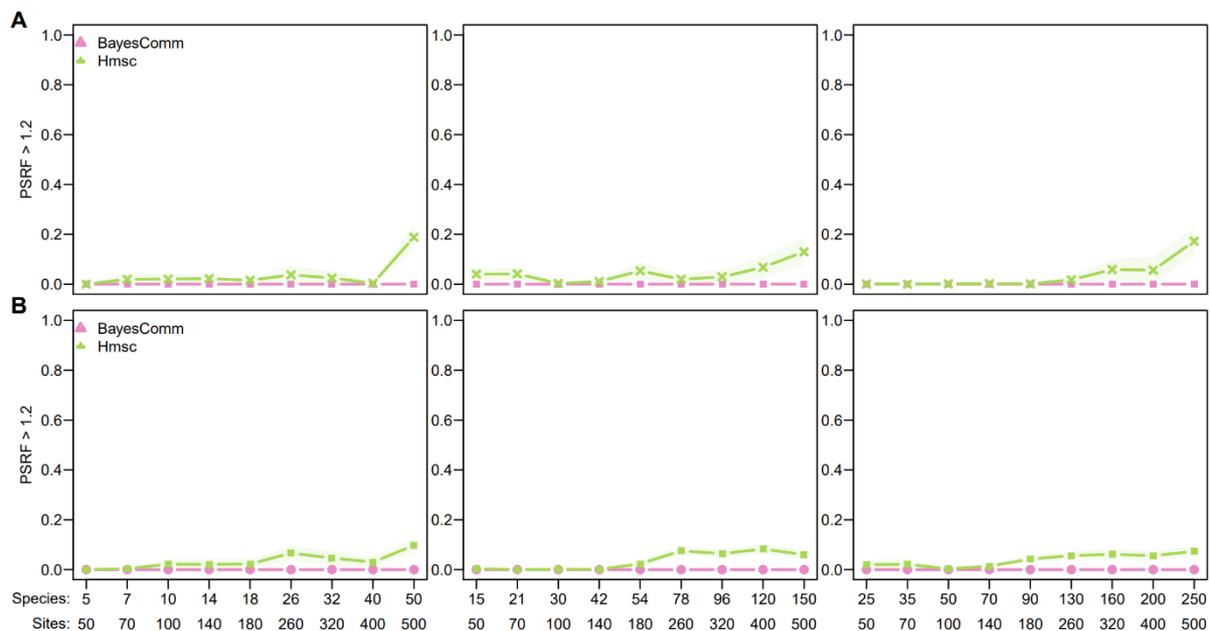

**Figure S5:** Rate of weights in percent with potential scale reduction factor > 1.2 with non-sparse and sparse association matrices in simulation scenarios for A) Hmsc (for factor loadings and beta estimates) and B) BayesComm (covariance and beta estimates).

## Covariance accuracy behavior

To further assess the jSDM's behavior in inferring the species-species association matrix, we set the number of species to 50 and increased the number of sites from 50 to 330. For each, step we computed the averaged (we sampled 5 scenarios for each setting) covariance accuracy and environmental RMSE.

BayesComm achieved at 330 sites around 0.82 accuracy and sjSDM around 0.80. sjSDM and BayesComm increased the covariance accuracy steadily with the number of sites, while Hmsc and gllvm stopped increasing their accuracy at around 0.68 accuracy (Fig. S6 A). sjSDM and BayesComm achieved in average 0.1 more accuracy than Hmsc and gllvm (Fig. S6 A).

All models achieved a similar RMSE over all scenarios (Fig. S6 B). sjSDM showed overall the highest RMSE (Fig. S6 B). All models decreased their RMSE with increasing number of sites (Fig. S6 B).

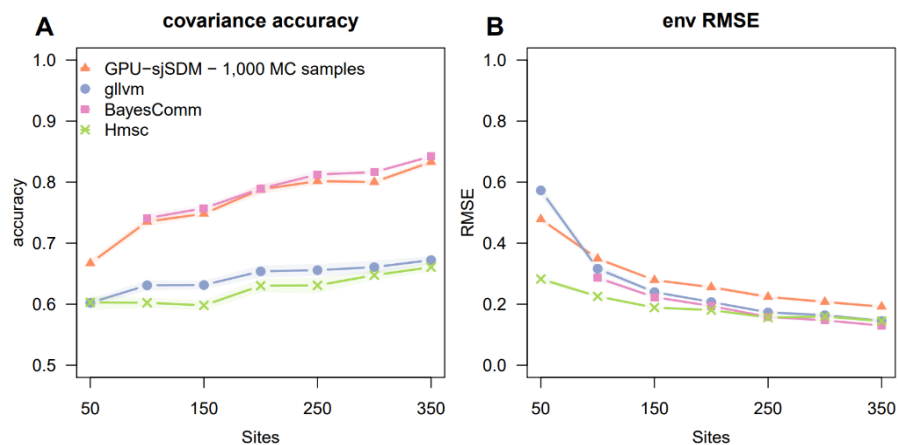

**Figure S6:** Results for examining the ability to recover the covariance structure as a function of the sites for G-sjSDM, BC, gllvm, and Hmsc. In the simulated species distribution scenarios, the number of species were constantly set to 50, but the number of sites were changed from 50 to 330 sites. A) Performance was measured by the accuracy of matching sings between estimated covariances matrices and true covariance matrices. B) Moreover, the root mean squared error for the environmental effects with the true coefficients were calculated.

## Simulation from a Latent Variable Model

We also simulated new data from a latent variable model varying the number of species from 10 to 100 (10, 50, and 100) and the number of latent variables (1 - 5) with a constant number of 200 sites. In all simulations, the species' environmental preference was described for two environmental covariates (beta).

Observed environment was sampled from a uniform distribution with a range of [-1, 1]. Unobserved environment (latent variables) was sampled from a normal distribution with mean equal 0.0 and standard deviation equal 1.0. For each scenario, we simulated 5 communities. The number of latent variables in gllvm was set to the real number of latent variables.

Environmental and latent coefficients (factor loadings) were sampled from a uniform distribution. We tested different scenarios: no latent variables at all (i.e. factor loadings equal zero), equal ranges/weighting between environmental coefficients and factor loadings (i.e. both were sampled from a uniform distribution within [-1, 1], 5:1 weighting of environmental coefficients to factor loadings (i.e. environmental coefficients were sampled from [-5, 5] and factor loadings from [-1, 1]), and a 1:5 weight of environmental coefficients to factor loadings (i.e. environmental coefficients were sampled from [-1, 1] and factors loadings from [-5, 5]).

We found for all three scenarios no large differences between the three JSDM (Fig. S7-S9). sjSDM was able to achieve the same or even better performance than the two LVM JSDM (Fig. S8A, C).

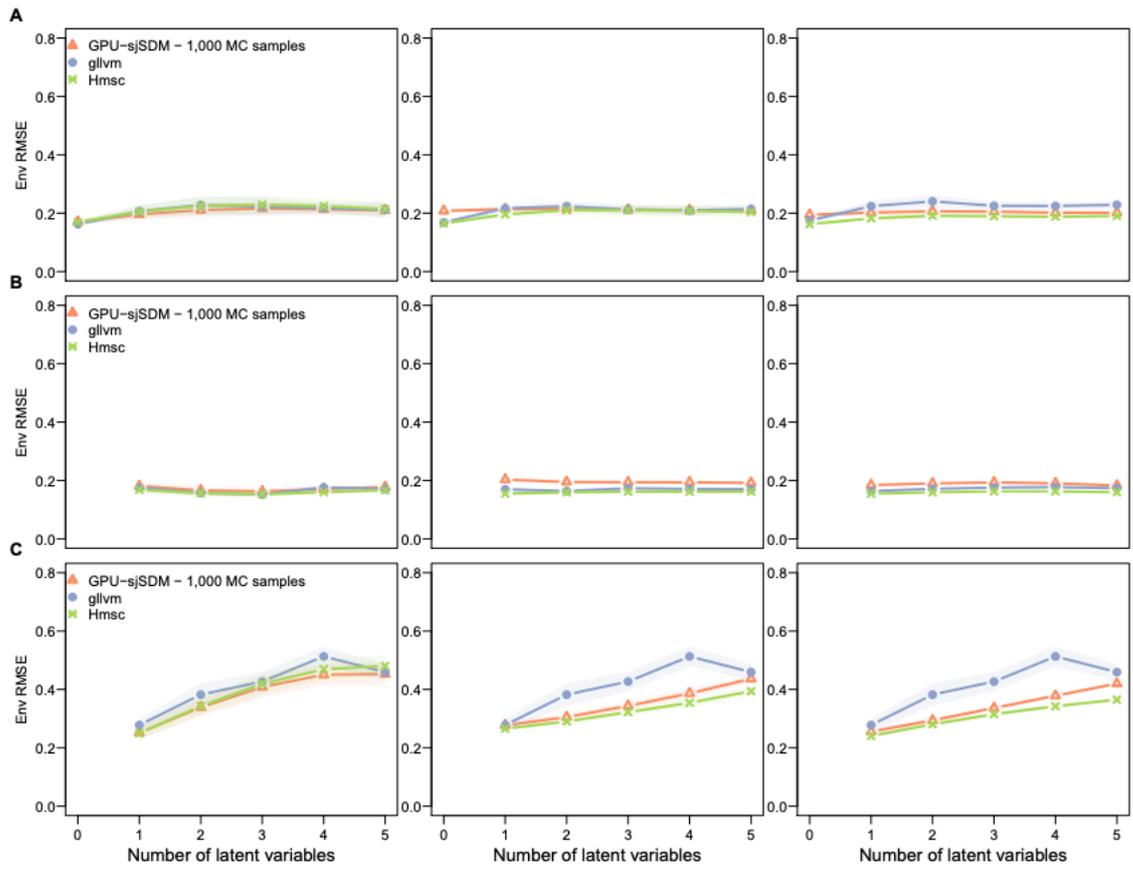

**Figure S7:** Results for communities simulated by the latent variable model. Models were fitted to different simulation scenarios: 10, 50 and 100 species with 1 - 5 latent variables. Number of environmental coefficients were set to two. For each scenario, the rmse error between true and estimated environmental coefficients over the five repetitions were averaged. A) shows the results for equal weighting B) for 5:1 (environment to latent) weighting and C) for 1:5 (environment to latent) weighting between the environmental and latent coefficients in the simulation.

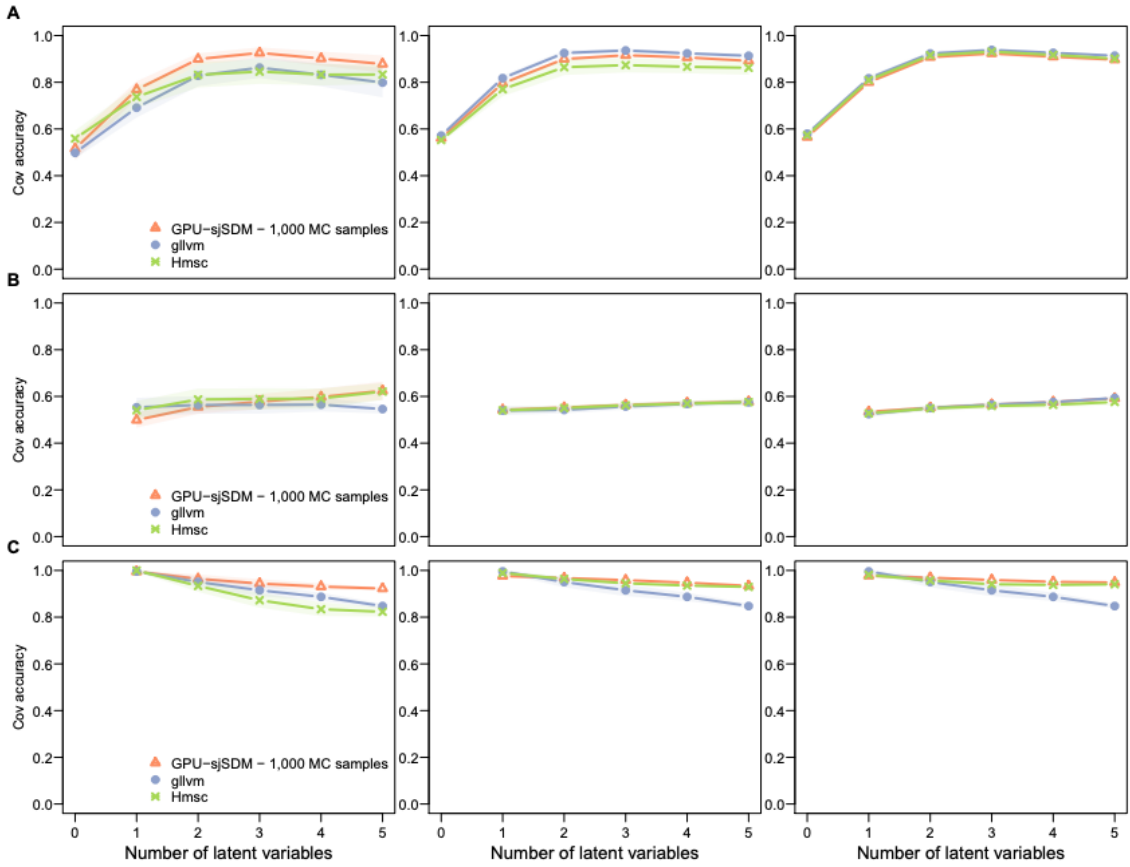

**Figure S8:** Results for communities simulated by the latent variable model. Models were fitted to different simulation scenarios: 10, 50 and 100 species with 1 - 5 latent variables. Number of environmental coefficients were set to two. For each scenario, the accuracy of matching signs of the covariances in the association matrix between true and estimated covariance matrix over the five repetitions were averaged. A) shows the results for equal weighting B) for 5:1 weighting and C) for 1:5 weighting between the environmental and latent coefficients in the simulation.

.

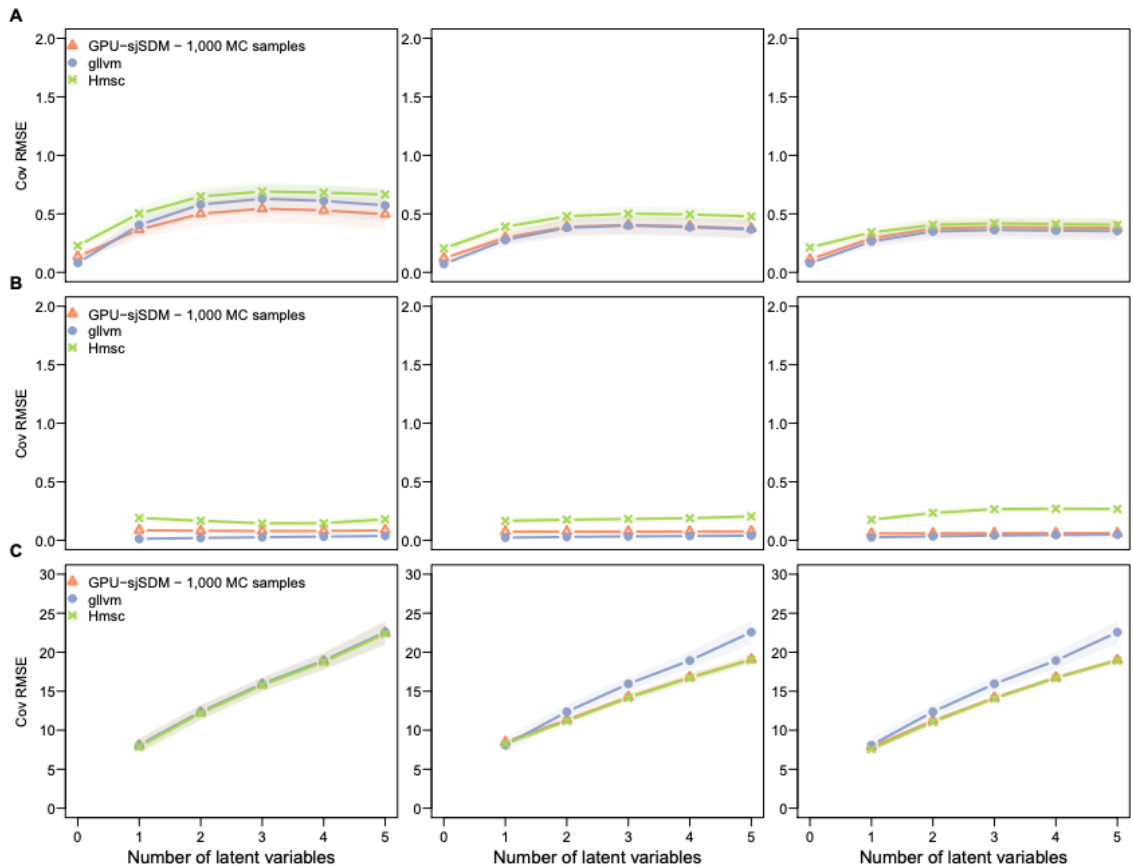

**Figure S9:** Results for communities simulated by the latent variable model. Models were fitted to different simulation scenarios: 10, 50 and 100 species with 1 - 5 latent variables. Number of environmental coefficients were set to two. For each scenario, the rmse between the true and estimated covariance matrix (normalized to correlation matrices) over the five repetitions were averaged. A) shows the results for equal weighting B) for 5:1 weighting and C) for 1:5 weighting between the environmental and latent coefficients in the simulation.

## Interpretation of the results for the eDNA dataset

For the eDNA dataset we found that the strongest negative associations are among the most abundant species while the strongest positive associations are between the rarest species (Fig. 5 A-B). Moreover, it appears that the most abundance species differ in their most import environmental coefficients (Fig. 5 D) suggesting that the abundant species occupy different niches within a community (site). Whether this pattern is caused by environmental filtering or biotic interactions (i.e. interspecific competition) is open for discussion (Kraft *et al.* 2015; Germain *et al.* 2018). But perhaps additional information such as traits or the comparison at different scales is required to disentangle the different factors (e.g. Boet *et al.* 2020). Also, we visualized here only the strongest 60 (30 negative and 30 positive associations) and we did not quantify the results. On the other hand, the rare species showed positive associations and more similar important environmental associations than the abundant species, which might suggest here a signal from biotic interactions rather than from environmental filtering. However, a more in-depth analysis is required to actually infer meaningful ecological insights from this dataset.

# Simulating from a process-based community model

To further explore the behavior of our new method compared to other JSDM implementations, we simulated communities from a process-based community model proposed by Leibold *et al.* 2020.

## A brief description of the process-based community model

A detailed explanation of the community model can be looked up in Leibold *et al.* 2020. The process-based model is based on spatially implicit site occupancy models. Colonization and extinction process describe the probability of occupancy of each species on each patch. Colonization depends on the number of immigrants, the environmental suitability, and the ecological interactions. Extinction depends on the environment and ecological interactions. Both processes are simulated over discrete time intervals. Leibold *et al.* 2020 proposed that the processes can be separated into three main contributions: space (immigration), environmental filtering, and co-occurrences of species (biotic interactions).

Additionally, Leibold *et al.* 2020 simulated key processes such as drift and dispersal. Each of the processes can be controlled.

## Simulation scenarios

We simulated the same 7 scenarios as Leibold *et al.* 2020. All scenarios had 12 species over 1000 sites over 5 discrete time intervals: A and B without ecological interactions but with a narrow and wide environmental niches, C and D with ecological interactions (interspecific competition effects) and narrow and wide environmental niches, E with half of the species (6) having ecological interactions, F without interactions but with 4 species having low dispersion, 4 species having medium dispersion, and 4 species having high dispersion, and G with different dispersions as in F but with interactions between the 4 species in the three groups (Table S5).

For each scenario, 5 temporal sequential realizations were simulated.

**Table S5:** Different scenarios for the simulations from the process-based community model from Leibold *et al.* 2020

| Scenario | Niche | Interactions | Dispersal |
|---|---|---|---|
| A | 0.8 | Colonization = extinction = 0.0 | 0.05 |
| B | 2.0 | Colonization = extinction = 0.0 | 0.05 |
| C | 0.8 | Colonization = extinction = 1.0 | 0.05 |
| D | 2.0 | Colonization = extinction = 1.0 | 0.05 |
| E | 0.8 | 6/12 species: Colonization = extinction = 0.0; | 0.05 |
|   |     | 6/12 species: Colonization = extinction = 1.0 | 0.05 |
| F | 0.8 | Colonization = extinction = 0.0 | 4/12 species: 0.01 |
|   |     |   | 4/12 species: 0.05 |
|   |     |   | 4/12 species: 0.1 |
| G | 0.8 | Colonization = extinction = 1.0 | 4/12 species: 0.01 |
|   |     |   | 4/12 species: 0.05 |
|   |     |   | 4/12 species: 0.1 |

## Model estimation

Following Leibold *et al.* 2020, we fitted the environmental variable E as main effect and as quadratic effect. Also, 50 spatial eigenvectors were fitted as main effects to account for space. We fitted sjSDM, Hmsc, BayesComm, and gllvm on each time step (5 discrete time intervals) for each scenario. The exact parameters of the models are described in Table S6.

For each model, the covariance matrix was extracted and if necessary normalized to [-1,1].

**Table S6:** Model settings of JSDM implementations Hmsc, BayesComm, gllvm, and sjSDM for process-based community simulations.

| Model | Parameter | Value |
|---|---|---|
| Hmsc | Iterations (MCMC samples) | 50,000 |
| | Burnin | 5,000 |
| | Thinning | 50 |
| | Number of chains | 1 |
| | Distribution | Binomial with probit link |
| BayesComm | Iterations (MCMC samples) | 50,000 |
| | Burnin | 5,000 |
| | Thinning | 50 |
| | Number of chains | 1 |
| | Distribution | Binomial with multivariate probit link |
| gllvm | Number of latent variables | 2 |
| | Distribution | Binomial with probit link |
| | Starting values | 'zero' (residulas). If model did not converge, retry with 'random' and if model still didn't converge, retry with 'res' |
| sjSDM | Learning rate | 0.1 |
| | Iterations | 100 |
| | MC-samples for each species | 1,000 (GPU) |
| | Distribution | Binomial with multivariate logit link |

## Evaluation of the JSDM for the process-based simulations

While Leibold *et al.* 2020 was more interested in investigating the ability of JSDM to separate the three functional processes of the simulation (space, biotic, and abiotic effects), we were mainly interested in the association structures found by the JSDM (3 of the 4 JSDM implementations are short of a method to separate the individual contributions. However, regarding the current criticisms of JSDM about the connection between associations and biotic interactions, we thought that comparing the estimated associations might be of highest interested.

The extracted (and normalized) covariance matrices were summed up over the 5 temporal steps and divided by 5. Previous the summarization, we took the absolute values of the covariance matrices since the simulation consisted of asymmetric associations whereas JSDM can only estimate symmetric associations and we were mainly interested in the question whether the models were able to find a signal in the associations – regardless of the sign of the covariance.

We then calculated the root mean squared error (RMSE) between the true association matrices (with entries of 1 between interacting species) and the estimated association matrices from the JSDM.

## Results of the JSDM for the process-based simulations

With no interactions (scenarios A, B, and F), we found that sjSDM, BayesComm and gllvm estimated associations close to zero (sjSDM and BayesComm) or exact zero (gllvm). Hmsc, however, did not estimate associations to zero (Fig. S10 A, B, F). With interactions (scenarios C - E, G), sjSDM and BayesComm estimated similar association structures as in the true association matrices (Table S5, red rectangles Fig. S10 C - E, G) whereas Hmsc estimated overall higher associations between all species, but with higher values for the positions of the true associations. Generally, sjSDM and BayesComm achieved the lowest RMSE in all scenarios except for gllvm, which Gllvm estimated associations close to zero in all scenarios (Fig. S10).

Overall, the LVM based JSDM (Hmsc and gllvm) estimated either too many associations unequal to zero (Hmsc) or all associations as or close to zero (gllvm). The difference between Hmsc and gllvm might be attributed to the difference of the estimation method, while gllvm uses variational inference or laplace approximation and Hmsc MCMC sampling, both methods are Bayesian, and the results indicate a problem in the prior for the latent variables (or factor loadings): The prior might be too restrictive for gllvm and not restrictive enough for Hmsc. Nevertheless, Hmsc showed again the typical block structures in the estimated covariance matrix (Fig. S10G) which is typically for low-rank approximations.

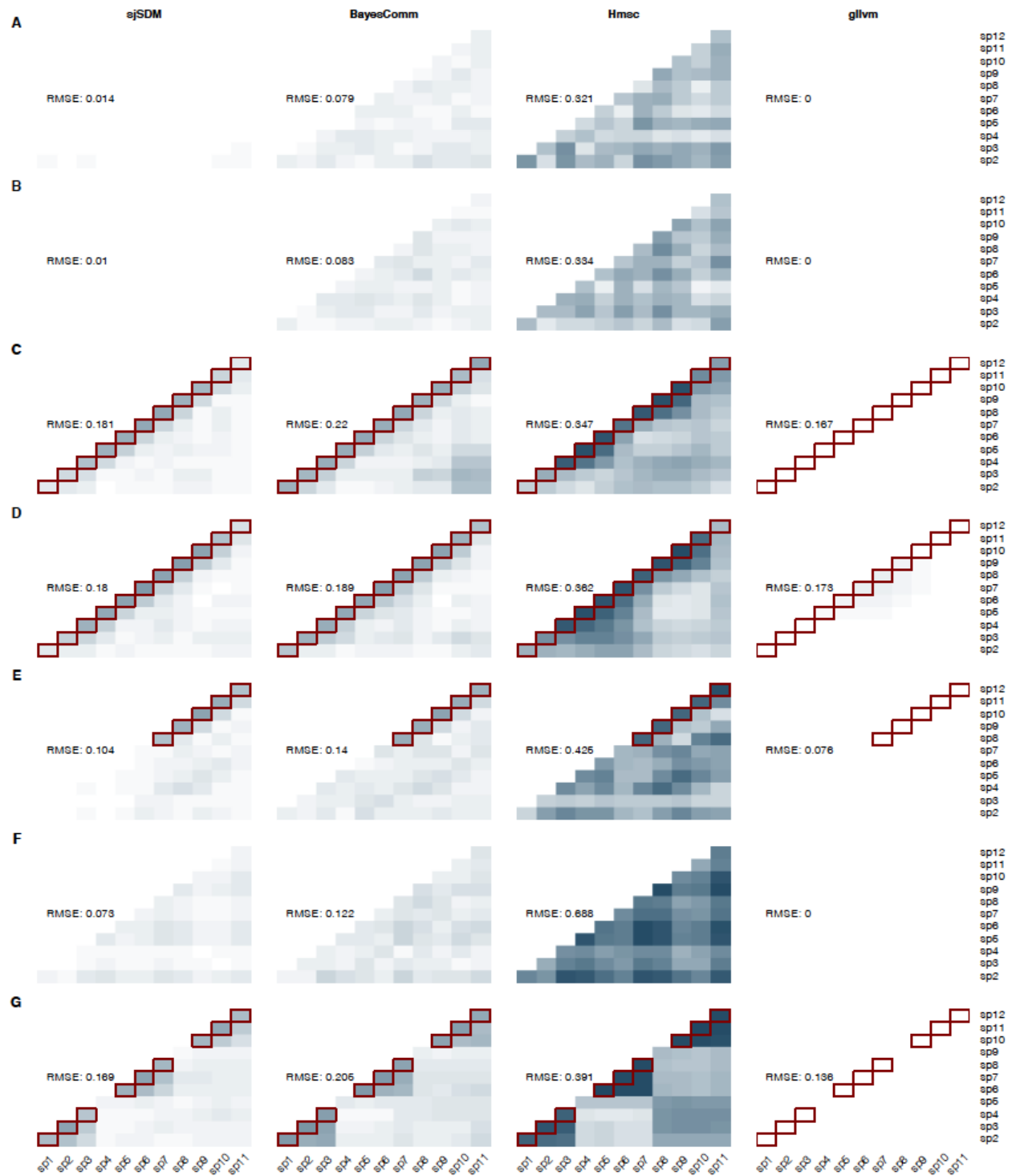

**Figure S10:** Results for simulations by the process-based community model from Leibold *et al.* 2020. Following Leibold *et al.* 2020, we tested 7 different scenarios with different niche widths (A-D), with and without interactions (A, B, and F without interactions), and different dispersal rates (F and G). Red rectangles show the true association matrices. For each scenario, 5 communities were temporally sequentially sampled and each of the JSDM were fit

to all 5 realizations. Afterwards, the normalized ([-1,1]) covariance matrices were averaged over their absolute values and the root mean squared errors to the true association matrices were calculated.